\documentclass[reprint]{JASAnewSamiya}

\usepackage{footmisc}

\usepackage{amsmath}
\usepackage{amssymb}
\usepackage{hyperref}
\usepackage{graphicx}

\newcommand{\w}{\omega}

\newcommand{\B}{\beta}

\newcommand{\expn}[1]{e^{#1}}

\renewcommand{\r}{\rho}

\newcommand{\Ap}{A_p}
\newcommand{\bp}{b_p}
\newcommand{\Bu}{B_u}

\newcommand{\so}{s_o}

\newcommand{\wmax}{\w_{max}}
\newcommand{\vstapes}{v_{stapes}}
\newcommand{\pconj}{\overline{p}}

\renewcommand{\xi}{x_i}

\newcommand{\stilde}{\tilde{s}}

\newcommand{\kx}{k}

\newcommand{\kB}{k_{\B}}

\renewcommand{\Re}{\text{Re}}
\renewcommand{\Im}{\text{Im}}

\newcommand{\aftertbme}[1]{ #1}

\newcommand{\aftertbmeii}[1]{#1}

\usepackage{mathtools}

\usepackage{enumitem}
\setitemize{noitemsep,topsep=0pt,parsep=0pt,partopsep=0pt}

\usepackage{xcolor}
\usepackage{tcolorbox}

\tcbset{width=0.9\textwidth,boxrule=0pt,colback=red,arc=0pt,auto outer arc,left=0pt,right=0pt,boxsep=5pt}

\renewcommand\thempfootnote{\arabic{mpfootnote}} 

\begin{document}

\title[JASA/An Analytic Physically Motivated Model of the Mammalian Cochlea]{An Analytic Physically Motivated Model of the Mammalian Cochlea}
\author{Samiya A Alkhairy}
\affiliation{Massachusetts Institute of Technology, Cambridge, MA, 02139, USA}
\email{samiya@alum.mit.edu, samiya@mit.edu}

\author{Christopher A Shera}
\affiliation{University of South California, Los Angeles, CA, 90033, USA}

\preprint{Author, JASA}		

\date{\today}

\begin{abstract}


We develop an analytic model of the mammalian cochlea. We use a mixed physical-phenomenological approach by utilizing existing work on the physics of classical box-representations of the cochlea, and behavior of recent data-derived wavenumber estimates. Spatial variation is incorporated through a single independent variable that combines space and frequency. We arrive at closed-form expressions for the organ of Corti velocity, its impedance, the pressure difference across the organ of Corti, and its wavenumber. We perform model tests using real and imaginary parts of chinchilla data from multiple locations and for multiple variables. The model also predicts impedances that are qualitatively consistent with current literature. \added{For implementation, the} model can leverage existing efforts for both filter bank and filter cascade models that target improved algorithmic or analog circuit efficiencies. The simplicity of the cochlear model, its small number of model constants, its ability to capture the variation of tuning, its closed-form expressions for physically-interrelated variables, and the form of these expressions that allows for easily determining one variable from another make the model appropriate for analytic and digital auditory filter implementations as discussed here, as well as for extracting macromechanical insights regarding how the cochlea works.

\end{abstract}

\maketitle


\section{Introduction}
\label{s:intro}

\subsection{Scope}
In this paper we present a model of the mammalian cochlea. \replaced{that}{The developed model} can be used for designing auditory filters for machine hearing (cochlear implants and hearing aids), compressing audio files, speech processing, and speech recognition. In addition to these engineering applications, the model can be used to help understand how the cochlea works; such as determining properties of power amplification and absorption that underlie cochlear responses, which is of interest as high gain is a distinctive feature of the normal cochlea. \added{It is possible to use the model for these applications because it provides a single framework that has simple representations with desirable features for both mechanistic and response (or filter) variables.}

\added{In this paper, we develop and test the model as is appropriate for any of the aforementioned applications. We then discuss auditory filter design applications which directly illustrates the appropriateness of the model and the advantages of its features for engineering applications, and indirectly supports the model in general, and its use for other applications.}

\subsection{Literature Survey}


\deleted{In this paper, we develop the model as is appropriate for any of the aforementioned applications, then focus our discussion on auditory filter design applications. It is relevant to provide a general  survey of auditory filter models and mechanistic models as current models in both categories have properties that are desirable even in the context of designing auditory filters alone.}

\subsubsection{Current Cochlear Model: Auditory Filters or Mechanistic Models}


Most other models of the cochlea fall \added{either} into the category of auditory filters \textit{or} the category of mechanistic models, as we shall describe here \cite{saremi} \footnote{While we make a distinction between auditory filter models and mechanistic models based on current literature, certain applications lie on the continuum between these two purposes: such as determining mechanisms (e.g. variation of gain) that underlie masking phenomenon in higher central nervous system studies; and determining what mechanistic differences underlie functional response differences (e.g. skewness of frequency response curves) between the base and the apex. For such applications, it is desirable to use a model, such as ours, that has properties of both categories of models and has a capacity to bridge between these two categories. }. \added{It is relevant to provide a brief survey of both the auditory filter models and mechanistic models for those readers interested in applications - including those purely interested in only either the engineering or scientific applications. This is because while certain model features are necessary only for a particular application, they are quite desirable for another application - whether in terms of increasing intuition, easing implementability or efficiency, and introducing desirable extensions of an application \added{\footnote{For example, having closed-form expressions for response / filter variables is highly desirable (almost necessary) for efficient auditory filter design. This same feature, if in a model that can serve as a mechanistic model, can extend what is possible for scientific study applications - for example, it allows for determining power flux given measurable response data characteristics such as rise time of an impulse response. The benefit is further amplified given a simple closed-form expression with a small number of parameters}}.}

\added{\paragraph{Auditory Filters}} The primary criteria for current models purely for auditory filter design are accuracy and efficiency. These models have limited physical basis, and therefore cannot fully benefit in their design from information regarding the true cochlear system, such as the inherent spatial variation of bandwidth that is encoded in some mechanistic models of the cochlea such as \cite{zweig1991}. Nor can such models be used for estimating underlying physical variables and determining which properties of the model contribute towards cochlear functions or phenomenon such as masking. They are often constrained to single methods of implementation (filter cascade or filter bank, as described below), and either the time or the frequency domain. Examples of models purely for filter design include the pole gammatone filter (APGF) and the one zero gammatone filter (OZGF) \cite{Lyon96theall-pole,lyon}.

\added{\paragraph{Mechanistic Models}} On the other hand, current mechanistic models of the cochlea utilize and benefit from physics-based properties of the cochlea in their design and use these properties as the primary criteria in their development. This category includes parametric models - e.g. \cite{zweig, meaud, liu}, and nonparametric models \cite{deboer}. \added{Most parameteric mechanistic models generally assume a resonant simple harmonic oscillator at the foundation of the model \footnote{with fixed model constant values for the mass, spring and dashpot, chosen based on peak frequency and bandwidth and assuming pure resonance as the peak generating mechanism rather than taking the full model into consideration - which would naturally lead to different bandwidths and group delays than if purely generated by a resonant component} then include additional components on top of the resonant component}. Some of these mechanistic models contain representations for impedance and/or wavenumber \cite{zweig, deboer}. They do \textit{not} provide closed-form expressions for macromechanical responses such as pressure and velocity - closed-form expressions for response variables are desirable because of the simplicity of implementation for auditory filter design, ease of functional interpretations, and building intuition for macromechanical scientific study. Also, \deleted{for such complex mechanistic models,} it is not possible, practically, to systematically estimates model constants that result in desirable bandwidths and quality factors \added{for these mechanistic models}. Therefore, it is difficult for such mechanistic models to be used for filter design \added{\footnote{Some refer to the response variables of mechanistic models as auditory filters.  This does \textit{not} mean that those models are appropriate for auditory filter design applications.}} \added{\footnote{Note that a few papers that fall under the category of mechanistic models suggest extending their models towards auditory filter design applications. However, to our knowledge, all such instances cannot plausibly satisfy key features required for auditory filter design - for example, there are no closed-form expressions for response / filter variables and the parameter values cannot be efficiently inverted from data \added[such statements are usually found as single sentence side notes in papers and likely caused by the insularity of mechanistic model and auditory filter camps - researchers in one camp generally do not have deep knowledge of the other]{ }}} or for inverting for the wavenumber and impedance and determining which aspects of the system are important for various filter functions or scientific phenomenon.


\added{Some model features necessary for one of the above two applications are quite desirable for the other application. Therefore, a single framework that takes this into consideration is quite desirable regardless of the reader’s application of interest. Unlike most models, the proposed model does not fall solely under one of the above categories. As it contains components and desirable features of both categories, it can be used for both filter and mechanistic applications as well as intermediate application.}



\subsubsection{Current Auditory Filters: Filter Banks or Filter Cascades}
\added{While we develop and test the model for any application - constrained by the model features, we mainly elaborate on its application for auditory filters in section \ref{s:discussion} to further support the model and both its engineering and scientific uses. Therefore it is relevant to introduce a survey of implementational aspects framed through the lens of auditory filters here.}

Most models of the cochlea that are used in designing auditory filters fall into one of two categories: (1) filter banks - e.g. \cite{zilany, sumner, patterson80, glasberg}, and (2) filter cascades - e.g. \cite{kates, lyoncascade, lyoncascadereview}. For either of these cases, an individual filter is a mathematical expression for cochlear responses and has $m$ model constants. A \textit{set} of $n$ of these filters is required for processing a general signal. Filter banks can be thought of as set of filters in parallel, and filter cascades a set of filters in series \added{\footnote{Note that some authors refer to both parallel and cascade configurations both as filterbanks. Whereas we, amongst others, refer to parallel configurations as filter banks and series configurations as cascade filters}}. Each filter is centered around a particular peak frequency, $\text{CF}(x)$, and parameterized in such a way that the modeler must determine $m$x$n$ values \added{\footnote{unless there is a function relating parameter values across filters}}. \added{Only in the case of filter bank models with a simple formulation and a very small number of parameters (e.g. second order / RLC band-pass filters), can the values of parameters can be easily chosen based on desired characteristics such as bandwidths.}


In the case of filter bank models, the response at a particular frequency is determined only by a single filter \footnote{as in the properties of the response at a particular location only depend on the filter(s) associated with that location: some models include other components such as compression and low-pass filtering along with band-pass filtering, but we consider these to be parts of the same single filter.}. Such filter bank models do not \replaced[benefit might have been misleading word here, as cascade filters cannot benefit from cochlear physics in their derivation either]{benefit from}{parallel} the underlying cochlear physics. On the other hand, filter cascades are somewhat closer to the physics of the cochlea \added{\footnote{Note, however, that these filter models do \textit{not} use cochlear physics.  They do not benefit from guidance provided by cochlear physics in developing the filter expression - or in testing it or in choosing its parametric constraints. \added[One quick easy of supporting this statement is to see that if filter models did use cochlear physics, they would contain mechanistic variables representations as well]{ }}}: The response recorded at any point along the length of the cochlea is due to the properties at that point, as well as properties of points the signal encountered prior to it; similarly, in the case of filter cascades, the response at any particular frequency, is the cumulative result from multiple filters. \added{However, this comes at a cost - it is generally far more difficult to choose estimate parameter values for cascade filters than filter banks.}

\added{Unlike other models, the response variables of the proposed model can be expressed using either filter bank or filter cascade formulations. Hence, it can incorporate possible virtues of both formulations and benefit from implementations of either class.}


\subsection{Regarding our Model}
\label{s:regardingourmodel}

Our proposed model \added{is not constrained to be either an auditory filter or a mechanistic model, but rather contains variables required for both, and features desirable for related applications. The model} has \replaced{physically}{physical} underpinnings and supports a traveling wave - as seen is section \ref{s:physics}, which is important for scientific study and certain implementation schemes for auditory filters. In developing the model, we also utilize observed data and impose mathematical constraints as described in section \ref{s:phenomenon}. We use a combined space-frequency independent variable which underlies part of the variation of model responses along the length of the cochlea. In section \ref{s:model}, we provide closed-form analytic expressions for multiple variables relevant for one or both of the potential model applications: pressure differences across the organ of Corti, its wavenumber, organ of Corti impedance, and its velocity. These variables are related to each other through the underlying physics and their model expressions are parameterized by the same set of model constants. Therefore, given data corresponding to a \textit{single} variable, we can extract information regarding the motion and characteristics of the cochlea\added{, which is quite powerful}. We test the model using real and imaginary parts of data from the chinchilla using the wavenumber and velocity expressions.


After developing and testing the model, we focus our discussion in section \ref{s:discussion} on aspects of the model particularly relevant for one of its applications - auditory filter design. The model has certain similarities to \emph{both} filter banks and filter cascades, and hence can be implemented in a number of ways. Furthermore, these similarities allow us to leverage existing methods for digital and analog implementations that optimize computational efficiency and power use. We discuss approaches to implement the model for auditory filter applications, and exemplify one such approach \added{. We also give brief examples of how the model's features are desirable for different researchers.}. 

Our model for the mammalian cochlea has a small number of model constants \footnote{This is the case for any particular mammalian cochlea - with the exception of bats which function differently - in which the characteristic frequency map parameters are known and fixed.}: to first approximation, three, $\Ap, \Bu, \bp$, and the more refined version for the chinchilla, five, $a_{\Ap}, b_{\Ap}, a_{\Bu}, b_{\Bu}, \bp$. With only these model constants, the model can capture the variation of tuning along the entire length of the cochlea. The simplicity of the model and its small number of model constants allow for fast implementations and simple estimation of the model constants. In addition, these features make the model ideal for extracting insights regarding how it works (e.g. energy flow along the length of the cochlea).

It is important to note that while our model can be used for engineering and scientific purposes and bridge between them, the applications intended for our model do \emph{not} extend to studies of \textit{detailed} mechanics of the cochlea \added{\footnote{because our model does \textit{not} have separate representations for the various membranes, cells and fluids within the Organ of Corti}}. Those are better performed with complex mechanical models of the cochlea (e.g. \cite{meaud, liu}) which offer an ability to vary model constants (e.g. tectorial membrane properties) and observe resultant changes in responses - \emph{provided} an assumed set of values for model constants, and detailed structural properties. \added{Our model is also inappropriate, without further modification, for mechanistic studies that involve reverse traveling waves such as most otoacoustic emission studies - e.g. \cite{verhulst}, or applications that can only be achieved with nonlinearity - though suggestions for modification are included in section \ref{s:discussion}.}

\section{Physics Component of the Model}
\label{s:physics}

\subsection{Structure}

\begin{figure}[htbp!]
    \centering
    \includegraphics[scale = 0.33]{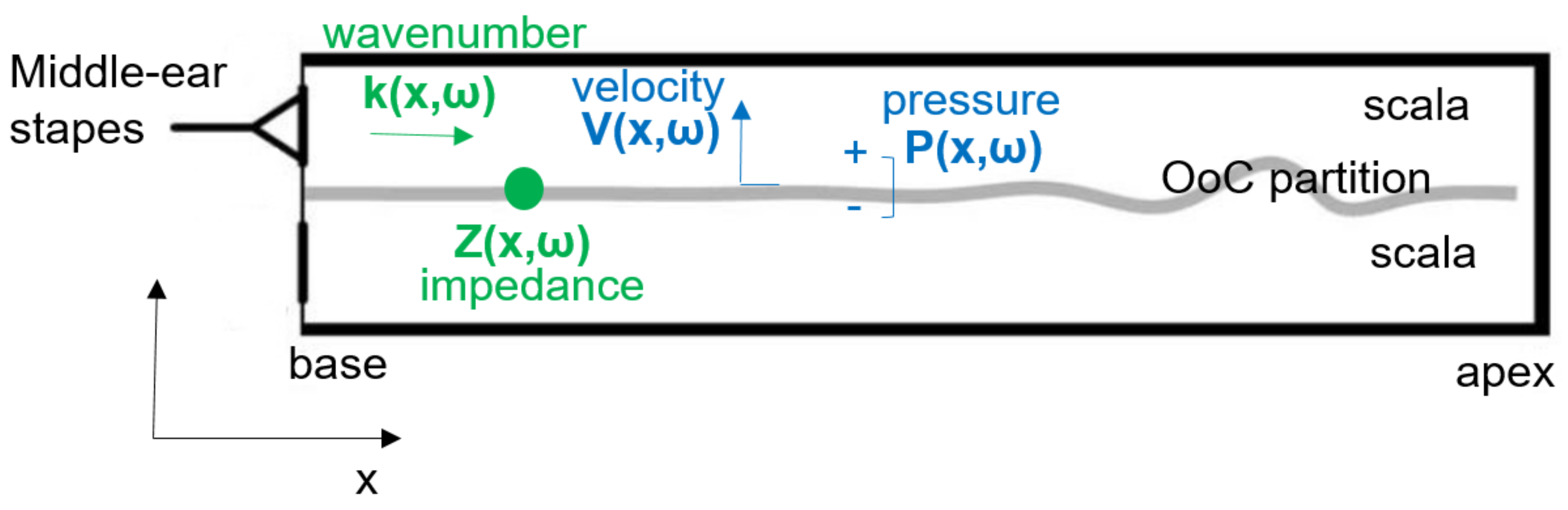}
    \caption[Box representation]{The figure (color online) illustrates the classical box representation of the cochlea. The scalae fluid compartments are separated by a single partition (with unknown properties) representing the OoC (in gray). We have annotated the figure with the transverse and longitudinal dimensions, as well as the response variables for transpartition pressure $P$ (pressure difference right across the OoC partition), and the partition velocity, $V$, which we have chosen to be positive in the upwards direction. The stimulus to the cochlea is specified by the stapes of the middle ear.  The figure is modified from \cite{sherawavenumber}. }
    \label{fig:box}
\end{figure}

We assume an uncoiled cochlea \cite{steelecoil} and choose a coordinate system such that the cochlea extends longitudinally in $x$ from the base to the apex.  We assume a classical box representation for the cochlea which consists of two fluid filled compartments, called the scalae, separated by the organ of Corti (OoC) which can be treated as a single partition \footnote{\aftertbmeii{Note that the model is based on macromechanical physics and hence there is no explicit representation for the basilar membrane in the model - the partition that represents the Organ of Corti includes its multiple membranes, cells, and fluid spaces.}}.
We describe the structure, variables, input and notation for the box model in Fig. \ref{fig:box}. For the derivation of the classical box representations and associated assumptions, we refer the reader to \cite{watts,neely}.

\subsection{Motion}

In response to an acoustic signal, the stapes vibrates and launches a pressure difference wave that propagates along the length of the cochlea from the base in the direction of the apex. The pressure difference wave interacts with the nonrigid OoC partition which has properties that depend on space and frequency. We assume linearity, which is presumed to be valid at low sound intensities \cite{geisler} in order to develop the model in the frequency domain. Additionally, the cochlear amplifier, which is the subject of much interest to the scientific community, primarily functions at low stimulus levels.

We use the notation $P$ for the short-wave pressure difference between the two scalae right across the OoC partition \footnote{\aftertbmeii{This short-wave approximation is derived from the two-dimensional box model of the cochlea, which is generally assumed to be an appropriate simple approximation for the three-dimensional model.}}. The short-wave approximation is valid where the wavelength is small relative to the scalae height, and is an appropriate assumption close to the peak region of the pressure wave. We are particularly interested in the peak region as it holds the most important information that is transmitted to the brain and is important for how we hear sounds. We assume (for simplicity and because the focus is on the peak region) that the pressure is short-wave everywhere - see appendix \ref{s:choiceofBC} which illustrates that, to first approximation, the long-wave effect can be neglected.

This pressure propagation in the forward direction can approximately be described by the following equation, 
\begin{equation}
    \frac{dP}{dx}(x, \w) + i\kx(x, \w) P(x, \w) = 0 \;.
    \label{eq:kxPclassical}
\end{equation}
The main point of this equation is that the pressure is simply a traveling wave, and its properties vary with space and frequency based on the wavenumber, $\kx$. The wavenumber specifies how the pressure wave changes as it propagates along the length of the cochlea. The dependence of the wavenumber on space and frequency is such that the pressure wave peaks in the base for high frequencies and in the apex for low frequencies. As we assume linearity, the pressure is directly proportional to the stapes velocity - for more details regarding how we treat the basal boundary condition, see appendix \ref{s:choiceofBC}.

The pressure wave in the scalae and the OoC partition properties both influence each other: (1) The effective OoC impedance, $Z$ partially determines the pressure wavenumber; In the region of the peak, which is what we are most interested in, this impedance is related to the wavenumber through the following equation, which depends on the density of the fluid and frequency,
\begin{equation}
    Z(x, \w) = \frac{-2i\r\w}{\kx(x, \w)}\;.
    \label{eq:kxZclassical}
\end{equation}

In addition, (2) the pressure difference across the OoC causes it to vibrate with velocity, $V$, according to,
\begin{equation}
    V(x, \w) = \frac{P(x, \w)}{Z(x, \w)} \;.
    \label{eq:Zclassical}
\end{equation}

As mentioned earlier, such equations have been previously derived - e.g. see \cite{watts,neely}. \aftertbmeii{Unlike early classical models, we do \textit{not} assume a particular structure for the OoC, and hence the variables $k, Z, P, V$ are unknown at this stage.} 

\subsection{Scaling symmetry}

Empirically, velocity responses have been shown to approximately be functions of a \textit{normalized} frequency \cite{bekesy,zweig76}, $\B$,
\begin{equation}
\B(x, \w) \triangleq \frac{f}{\text{CF}(x)} \;,
\end{equation}
in terms of $\text{CF}(x)$, which is the characteristic frequency, or peak frequency at a particular location $x$. The value of $\text{CF}(x)$ is known for many species, including chinchilla and humans \cite{greenwood, muller,liberman}.


To simplify model development, we assume that the wavenumber can be described sufficiently well by assuming, $\kx = \kx(\B)$. We shall refer to this assumption as scaling symmetry of the wavenumber. Note that we use the same notation for the wavenumber we introduced earlier despite changing the independence variable for simplicity. 



\section{Phenomenological Component of the Model}
\label{s:phenomenon}

We construct an expression for the wavenumber, $\kx$, based on phenomenon and constraints detailed in this section, then use it along with the physical inter-relations (Eqs. \ref{eq:kxPclassical} - \ref{eq:kxZclassical}) to determine the model expression for velocity and other variables. Our approach is therefore a mixed physical-phenomenological approach. 

\begin{figure}[htbp!]
    \centering
    \includegraphics[scale = 0.9]{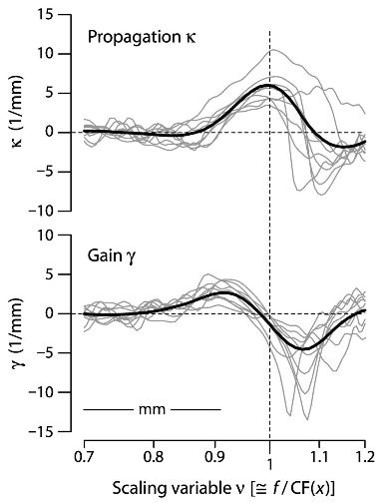}
    \caption[Wavenumbers derived from data]{Loess trends of real ($\kappa = \Re\{\kx\}$) and imaginary ($\gamma = \Im\{\kx\}$) parts of the estimated wavenumber of \cite{sherawavenumber} are in black, and individual curves from locations with CF$=8-10$ kHz are in grey. The individual curves were estimated from Wiener Kernel data from auditory nerve fibers of chinchilla. The wavenumber parts are shown as a function of the scaling variable, which is equivalent to our $\B$ in the basal half of the cochlea. Notice that the zero crossing of $\gamma$ and the maximum of $\kappa$ occur at approximately $\B = 1$. The figure is from \cite{sherawavenumber}. The  black curve (the trend line) captures the general shape and we use it to guide us in constructing an expression for the wavenumber. Curves from the apex follow a similar trend but differ quantitatively.}
    \label{fig:wavenumberdata}
\end{figure}

To construct the wavenumber, $\kx$, we utilize published observations derived from chinchilla data, which shows the estimated real and imaginary parts of the wavenumber in Fig. \ref{fig:wavenumberdata}, as a function of the normalized frequency, $\B$ \cite{sherawavenumber}. 

In addition to the general features derived from data, we assume that the general features of the wavenumbers are qualitatively consistent across species and regions we are interested in such as the base and apex. We construct an expression for the wavenumber with these in mind and such that the wavenumber expression is closed-form, and its integral (for velocity) is closed-form. We also impose the following constraints, which will lead to an intentional discrepancy \replaced{beween}{between} the behavior of our model wave number and that of data from \cite{sherawavenumber}.

\begin{itemize}
    \item Constraint 1: The model $\Re\{\kx\}$ is positive everywhere to have a purely forward traveling pressure wave. This constraint should be understood in the context that $\kx = \kx(\B)$\footnote{The constraint that $\kx = \kx(\B)$ ties together the space domain and frequency domain. In other words, it ties together the traveling wave perspective (which is the native domain for explaining and studying how the cochlea works), and the transfer function perspective (which is the native domain for functional aspects of cochlear responses and is the domain in which most data - including Wiener Kernel data - is collected).}.
    \item Constraint 2: The model $\Im\{\kx\} > 0$ prior to the peak of the traveling wave, then $\Im\{\kx\} < 0$ beyond the peak in order that the amplitude of the pressure traveling wave monotonically increases then monotonically decreases. This constraint is due to data of the pressure at a single location across frequency \cite{dong}, and an assumption that $P=P(\B)$ which is presumed to hold except very close to the stapes \cite{thesis}.
\end{itemize}


\section{Model Developments and Tests}
\label{s:model}

To develop our model, we first construct an expression for the wavenumber, then derive expressions for the remaining variables. As we present our expressions and derivation, we test the model qualitatively and quantitatively. Note that a model test on any single variable is a test of the entire model due to the physical relationship between the variables.

\subsection{Wavenumber}
\label{s:wavenumber}

The model expression we construct for the wavenumber, $\kx$, is as follows \footnote{Note that we defined this normalized dimensionless wavenumber, $\kB = \kx \frac{l}{\B}$ for simplicity of model construction. We relate $k_\beta$, the wavenumber with respect to $\B$, and $\kx$, the wavenumber with respect to $x$ (perhaps better denoted as $k_x$) through $k_\beta d\beta = \kx dx$ - note that $\B$ can be thought of as transformed $x$. This also allows us to rewrite Eq. \ref{eq:kxPclassical} as $\frac{dP}{d\B}+i\kB P = 0$ which simplifies deriving closed-form expressions specially for the \textit{transfer function} perspective of cochlear responses with appropriate basal boundary condition assumptions explained in appendix \ref{s:choiceofBC}.},
\begin{equation}
\kx \frac{l}{\B} = 2\Bu \frac{i\B+ \Ap}{(i\B-p)(i\B-\pconj)}\;.
\label{eq:kxfull}
\end{equation}
For compactness, and defining poles and zeros, we shall define an independent variable, $s$,
\begin{equation}
    s \triangleq i\B \;.
\end{equation}
The expression for the normalized wavenumber is a rational transfer function that has a pair of complex conjugate poles, $p = i\bp - \Ap$ and  $\pconj = -i\bp - \Ap$, as well as a real zero at $s = -\Ap$. The three model constants, $\Ap, \bp, \Bu$ take on positive real values. The constant, $l$ is the space constant of the cochlear map, $\text{CF}(x)= f_{max}\expn{-\frac{x}{l}}$, that is empirically known for a variety of species, including humans and chinchilla.

\added{Note that it is trivial to determine discrete forms of any of the variables’ expressions in this paper using transform methods - e.g. bilinear transform, impulse invariance, or pole-zero-mapping / matched Z transform methods. Hence, we do not include these expressions in this paper.}

\subsubsection{Note on the Form}

As can be seen from the expression above, the wavenumber satisfies desirable criteria: it is a closed-form expression, its integral is closed-form, it is in terms of $\B$ which couples its dependence on space and frequency, and is in terms of only three model constants. 

We note that a recent model \cite{zweig}, independently derived by Zweig, ended up with a wavenumber that is similar to our above expression in the sense of having a pair of complex conjugate poles, which is particularly encouraging for our model and his, as the models were developed using very different methods.

Our wavenumber expression can be written using the partial fraction expansion as,
\begin{equation}
\kx \frac{l}{\B} = \frac{\Bu}{s-p}
+ \frac{\Bu}{s-\pconj}  \xrightarrow{\text{sharp-filter approx.}} \frac{\Bu}{s-p} \;.
\label{eq:kxsharp}
\end{equation}
In the equation above, we have also provided expressions for the sharp-filter approximation of $\kx$, which holds when $|s-\pconj| \gg |s-p|$ or equivalently when $\Ap$ is small. Near the peak of $V$ (i.e. near $\B \approx \bp$), the sharp-filter condition is $\Ap \ll 2\bp$. The sharp-filter approximation provides useful intuition due to the simple nature of the expressions. 
 
\subsubsection{Information Regarding the Cochlea}

From the above expressions, we can simply determine closed-form expressions for the real and imaginary parts of $\kx$ and $Z$ when $\B$ is real. The closed-form expressions are simple enough to derive insights from. This is particularly relevant as $\Re\{\kx\}, \Im\{\kx\}, \Re\{Z\}, \Im\{Z\}$ encode certain aspects of how the cochlea works and are the subject of scientific interest. The real part of the wavenumber encodes aspects such as the wavelengths and phase and group velocities (how fast the pressure wave propagates along the length of the cochlea), as well as dispersivity. The imaginary part of the wavenumber encodes gain and dissipation. The real part of the effective impedance encodes effective positive or negative damping, and the imaginary part encodes effective stiffness. 

\subsubsection{Model Test}

To test the model, we fit our wavenumber expression to the data in Fig. \ref{fig:wavenumFit} according to the method of fitting in appendix \ref{s:fitmethod}. 
\begin{figure}[htbp!]
    \centering
    \includegraphics[scale = 0.33]{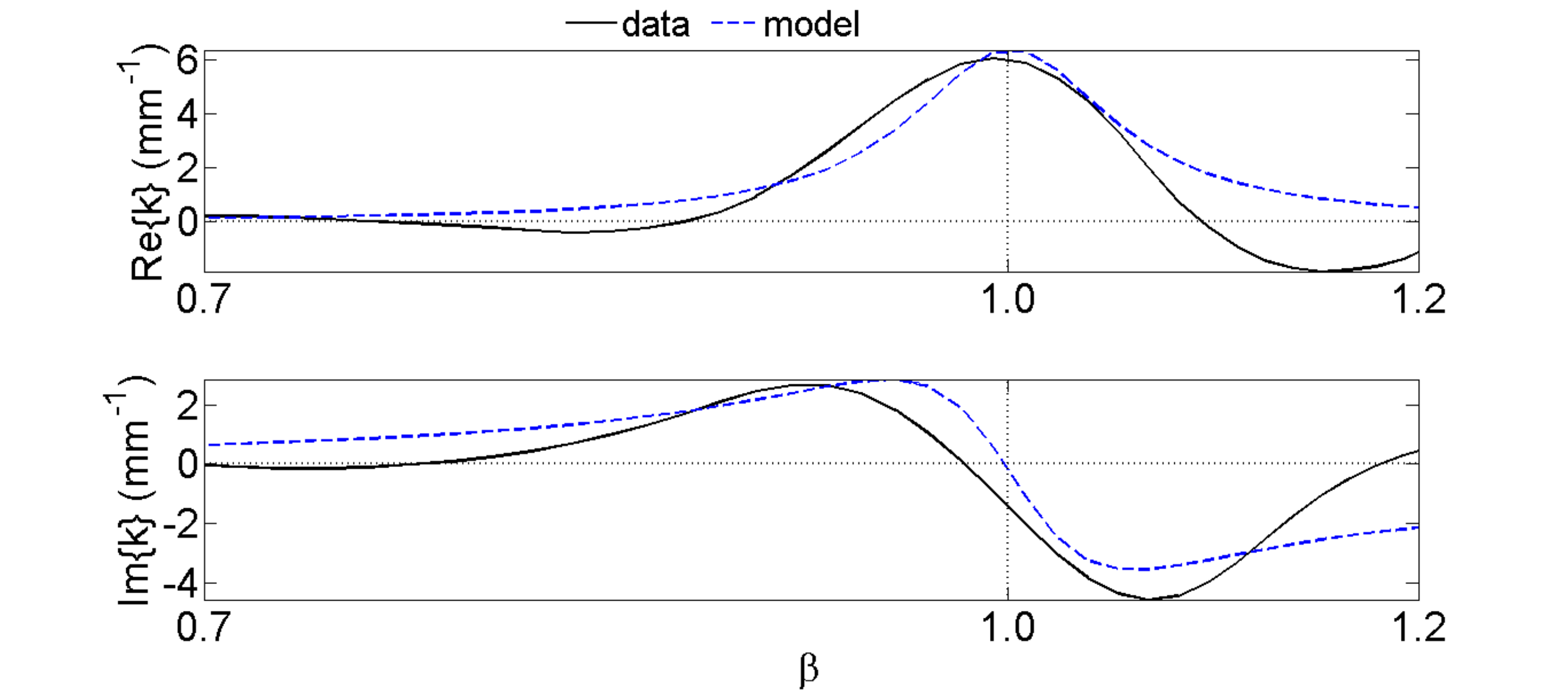}
    \caption[Model test using wavenumber]{Model test using wavenumber: The black line reproduces data from figure \ref{fig:wavenumberdata} which is from \cite{sherawavenumber}. The blue dashed line (color online) shows the model fits for chinchillas which have a cochlear map space constant $l = 3.8 $ mm (see \cite{sherawavenumber}). The top panel shows the real part of the wavenumber, which encodes propagation properties, and the bottom shows the imaginary part of the wavenumber which encodes gain properties. The $\Re\{\kx\}, \Im\{\kx\}$ are plotted as a function of the normalized frequency, $\B$. The best-fit model constants used to construct our model curves are: $\Ap = 0.05, \Bu = 1.3$, with $\bp = 1$ held as a fixed constant. }
    \label{fig:wavenumFit}
\end{figure}

Figure \ref{fig:wavenumFit} shows that the model has a peak in $\Re\{\kx\}$ that occurs when the normalized frequency $\B = 1$. This peak indicates that the local wavelength is shortest near the peak, and is a feature of the data that the model captures. The model deviates from the data in that  $\Re\{\kx\}$ in the model is always positive, as a result of our Constraint 1, whereas this is not the case with the data at higher values of $\B$.


The imaginary part of the wavenumber of both the model and data is first positive. The fact that $\Im\{\kx\} > 0$ here indicates that the pressure wave grows as it propagates along the length of the cochlea in this region. Then $\Im\{\kx\}$ has a zero crossing indicating that the pressure reaches its peak. Finally, $\Im\{\kx\}$ becomes negative indicating that the pressure decays.
There is a deviation between the model and the data for larger values of $\B$, where the model does not return to zero as quickly as the data. The model curve also does not become positive - a result of Constraint 2 which we imposed during model construction.

\subsection{Impedance}

We may use our phenomenological expression for the wavenumber, along with the physical equations, to determine a closed-form expression for the effective OoC impedance - a variable of special interest for understanding how the cochlea works from the macromechanical perspective. The impedance is of particular interest in model development as we can qualitatively compare our impedance with previous estimates in the literature to test our entire model.

The real and the imaginary parts of the impedance are expressed separately as,\begin{align}
    \Re\{\frac{Z}{2\pi \r l\text{CF}(x)}\} & =  \frac{\B}{\Bu} \frac{\B^2 + (\Ap^2 - \bp^2) }{\Ap^2 + \B^2}  \label{eq:impedanceparts} \\
    &  \xrightarrow{\text{sharp-filter approx.}} \frac{2}{\Bu} (\B - \bp) \nonumber \\
    \Im\{\frac{Z}{2\pi \r l\text{CF}(x)}\} & =  -\frac{\Ap}{\Bu} \frac{\B^2 + (\Ap^2 + \bp^2) }{\Ap^2 + \B^2}
    \nonumber \\    &  \xrightarrow{\text{sharp-filter approx.}} -\frac{2\Ap }{\Bu} 
    \nonumber\;.
\end{align}



\subsubsection{Notes on the Form}

Our $\Im\{Z\}$ is negative for all values of $(x,\w)$ indicating that the OoC is dominated by effective stiffness over effective mass and there is no `resonance' \footnote{The term effective is in the sense that it a sum of various properties of the OoC.}. Additionally, for the sharp-filter approximation, $\Im\{Z\}$ is constant with respect to frequency. Hence, we can assume linear elasticity for this part of the impedance, at least close to the peak of the traveling wave.

The real part of our effective impedance is negative at frequencies and locations prior to the peak of velocity or pressure, indicating negative effective damping (and hence net power amplification to the pressure wave). $\Re\{Z\}$ then becomes positive beyond the peak indicating positive effective damping (and hence net power absorption). For the sharp-filter approximation, $\Re\{Z\}$ is linear and has a zero crossing at $\B = \bp$.

\subsubsection{Model Test}

Recent parametric - e.g. \cite{zweig}, and nonparametric - e.g. \cite{deboer}, models of the impedance, and experimental estimates of the impedance -e.g. \cite{dong}, suggest that $\Im\{Z\} < 0$ prior to the peak in $V$ and at least shortly beyond it. This is in general qualitative agreement with our model. However, we must note that all of these estimates differ quantitatively, as well as qualitatively in terms of the dependence of their impedance on frequency and/or location.


The aforementioned other estimates of the impedance - e.g. \cite{deboer,zweig,dong}, have a real part that is negative then positive and crosses zero near the peak, which is qualitatively consistent with our model. However, the frequency at which this zero crossing occurs and the behavior away from this crossing varies across these estimates.

\subsection{Pressure and Velocity}
\label{s:velocity}
We use the phenomenological expression we constructed for the wavenumber, along with the physical inter-relations in order to derive the model expressions for pressure, $P$, 

\begin{equation}
 \frac{P(\B)}{\vstapes} = C  { \bigg( (s-p)(s-\pconj) \bigg)^{-\Bu}}  \;,
 \label{eq:P}
\end{equation}

and subsequently for velocity, $V$,

\begin{equation}
 \frac{V(x,\w)}{\vstapes} = C  \frac{i \Bu }{\rho l \text{CF}(x)} \frac{s+\Ap}{ \bigg( (s-p)(s-\pconj) \bigg)^{\Bu + 1}}  \;.
 \label{eq:velocity}
\end{equation}


The above equations are closed-form and in terms of the same set of three model constants, $\Ap, \bp, \Bu$, and in reference to a stapes velocity, $\vstapes$. The pressure and velocity are in reference to a constant, $C$ with units of pressure by velocity \footnote{Note that we have assumed the following regarding the integration constant from equation \ref{eq:kxPclassical}: its primary frequency dependence comes from the stapes velocity, which, for an impulse is a constant $\vstapes(\w) = 1$ (the pressure and velocity are directly proportional to $\vstapes(\w)$); the remaining frequency dependence due to proximity to the basal boundary is negligible for velocity except in the very base. Therefore, we may assume that $C$ is truly a (complex) constant, and independent of frequency. We shall use this assumption and presume it is valid except closest to the stapes \cite{thesis}. Indeed, we have studied the dependence of the normalized velocity response on two different boundary conditions for three locations along the length of the chinchilla cochlea, and found that the dependence on the choice of boundary condition does not have a major effect except closest to the stapes.}. The constant, $C$ is due to the effect of the region between the stapes and the short-wave and we treat it as an unknown constant. This is not an issue as the variable most relevant for auditory filter design is the response \textit{normalized} to its peak value at a particular location, $\mathcal{V}$, or $\mathcal{P}$. See appendix \ref{s:choiceofBC} for further details regarding the dependence of the response on the middle ear boundary condition. The normalized pressure and velocity are very similar near the peak for realistic model parameter values.

\subsubsection{Note on the Form}
\added{The forms for both response variables, $P, V$, show that they have peaks in magnitude \footnote{As a side note, but one that is relevant given likely passive-active differences, and the fact that $P$ is often neglected in studies: it can be easily derived from the $k-Z$ relationship that a model where the peak of velocity is generated due to a the zero-crossing of the imaginary part of the impedance (i.e. due to a resonance occurring at that point) would not have a peak in pressure magnitude there}. This is qualitatively consistent with measurements - e.g. \cite{dong}. For those interested purely in mechanistic studies, we note that, to our knowledge, no other models with mechanistic components have closed-form expressions for response variables \footnote{An exception is Helmholtz, but the underlying mechanism does not contain traveling waves, and has been shown to be an inappropriate model of the cochlea}. This extends the possibility of what can be scientifically studied using cochlear models and also what can be inferred from data about the model, and subsequently, the cochlea. Furthermore, the fact that the response variables (from a mechanistic standpoint) and the filter variables (from an engineering standpoint) are represented by the same \textit{closed-form expressions}  (for pressure and velocity) in our multivariate model, not only allows for the same simple model to be used by both engineers and scientists, but also provides a framework to bridge between them and an opportunity for them to overlap.} 

It is interesting to note that the pressure and velocity expressions are similar to a class of auditory filter models - the gammatone filter family \cite{Lyon96theall-pole,lyon5}: both have a pair of complex-conjugate poles raised to a power. \added{For those interested purely in engineering purposes,} this is encouraging and also quite useful since the gammatone filters are widely used, and they can be applied towards cochlear implants, hearing aids, and audio-engineering purposes. Hence, provided that the differences are taken into consideration, it is possible to use our model for the same applications as the gammatone filters. The similarities that may allow for leveraging current implementations are further discussed in section \ref{s:similaritiesOtherModels}.

\subsubsection{Model Test}

\begin{figure}[htbp!]
    \centering
    \includegraphics[scale = 0.12]{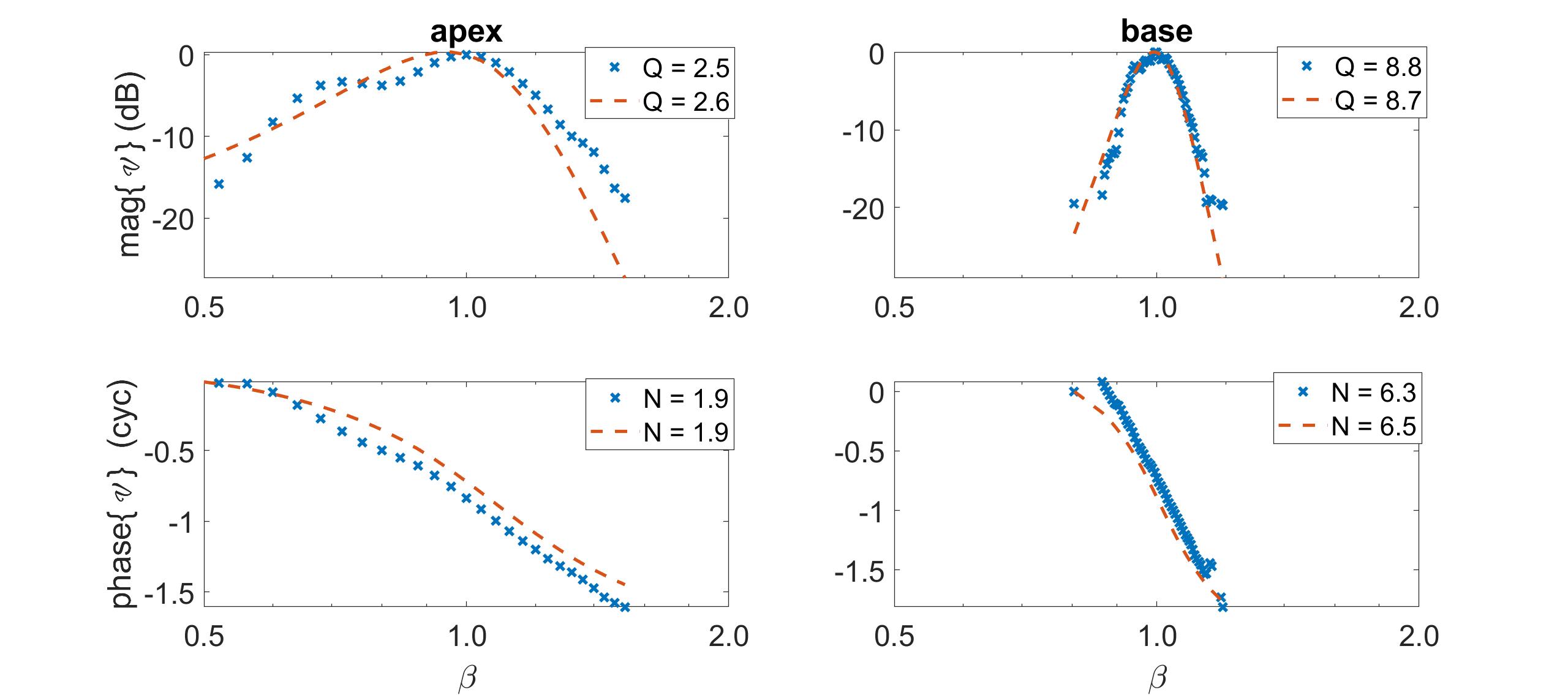}
    \caption[Model test using velocity expressions]{Model test using velocity expression: The model (red dashed - color online) expression for velocity response is fit to measured data (blue crosses) from Wiener Kernels of chinchilla neural data in the region of the peak \cite{ruggeroWK}. The fits are for data from the apex collected at a point where the characteristic frequency is 390 Hz (left) and for data from the base collected at a point where the characteristic frequency is 11 kHz (right). The magnitude (top) and phase (bottom) are plotted as a function of normalized frequency $\B$. The model constant $\bp$ is fixed at $\bp = 1$ for both fits. For the fit to the point in the apex, the estimated model constant values for $\Ap$, and $\Bu$ are 0.4, and 4.0 respectively, and the objective function value is 0.16. For the fit to the point in the base, the estimated model constant values for $\Ap$, and $\Bu$ are 0.15, and 5.0 respectively, and the objective function value is 0.09. The legends include computed values for the dimensionless quality factor, $Q$, derived from the equivalent rectangular bandwidth, and the normalized center frequency group delay, $N$ in cycles.}
    \label{fig:velocityNeural}
\end{figure}

To further test our model, we fit our velocity expression in Fig. \ref{fig:velocityNeural} to chinchilla auditory nerve fiber Wiener Kernel data \cite{ruggeroWK} using fitting methods described in appendix \ref{s:fitmethod}. While these neural Wiener Kernels are not direct measurements of OoC velocity, they have been shown to be a good approximation of the macromechanical responses in the peak region \cite{temchin} \footnote{Note that near the peak, the pressure (rather than admittance) dominates the velocity profile, and hence the choice of approximating the velocity or pressure with neural data does not make a significant difference. In addition, near the peak, the shape of the velocity profile varies more rapidly than $i\w$, and therefore choosing the velocity or displacement also does not make a significant difference.}. We fit to these data as they are available from multiple locations along the length of the cochlea - as opposed to mechanical measurements. 

We find that the model fits the data best near the peak of the magnitude curve. This is due to our focus on the peak during model construction, as well as our choice of objective function that we minimize - see appendix \ref{s:fitmethod}. The model magnitude curve has a quality factor \footnote{Defined as the dimensionless $Q \triangleq \frac{\text{CF}}{\text{ERB}}$, using the most commonly used construction for equivalent rectangular bandwidth, $\text{ERB} \triangleq \int_{f1}^{f2} |\mathcal{V}|^2 df$.  In computing $Q$, we have used all data points included in the plots. Note that points corresponding to lower magnitudes (further away from the peak) do not contribute much to $Q$.} similar to that of data \cite{sheratriangle}. The model also captures the slope of the phase data which is an important feature, and it has a normalized center frequency group delay \footnote{Defined as $N \triangleq -\frac{\text{CF}}{2\pi}\frac{d\phi}{df}\biggr\rvert_{f=\text{CF}}$ in periods of $\text{CF}$. In computing $N$, we averaged over 5 points around $\B=1$ to ameliorate problems associated with noisy data.} similar to that of data \cite{sheratriangle} \footnote{Note, however, that in this paper, we have defined the objective function to minimize the error in the responses and \emph{not} in $Q$ and $N$.}. Additional examples of our model fits to data are included in appendix \ref{s:morefits}.


\subsection{Model Constants}
\label{s:trends}

\begin{figure}[htbp!]
    \centering
    \includegraphics[scale = 0.33]{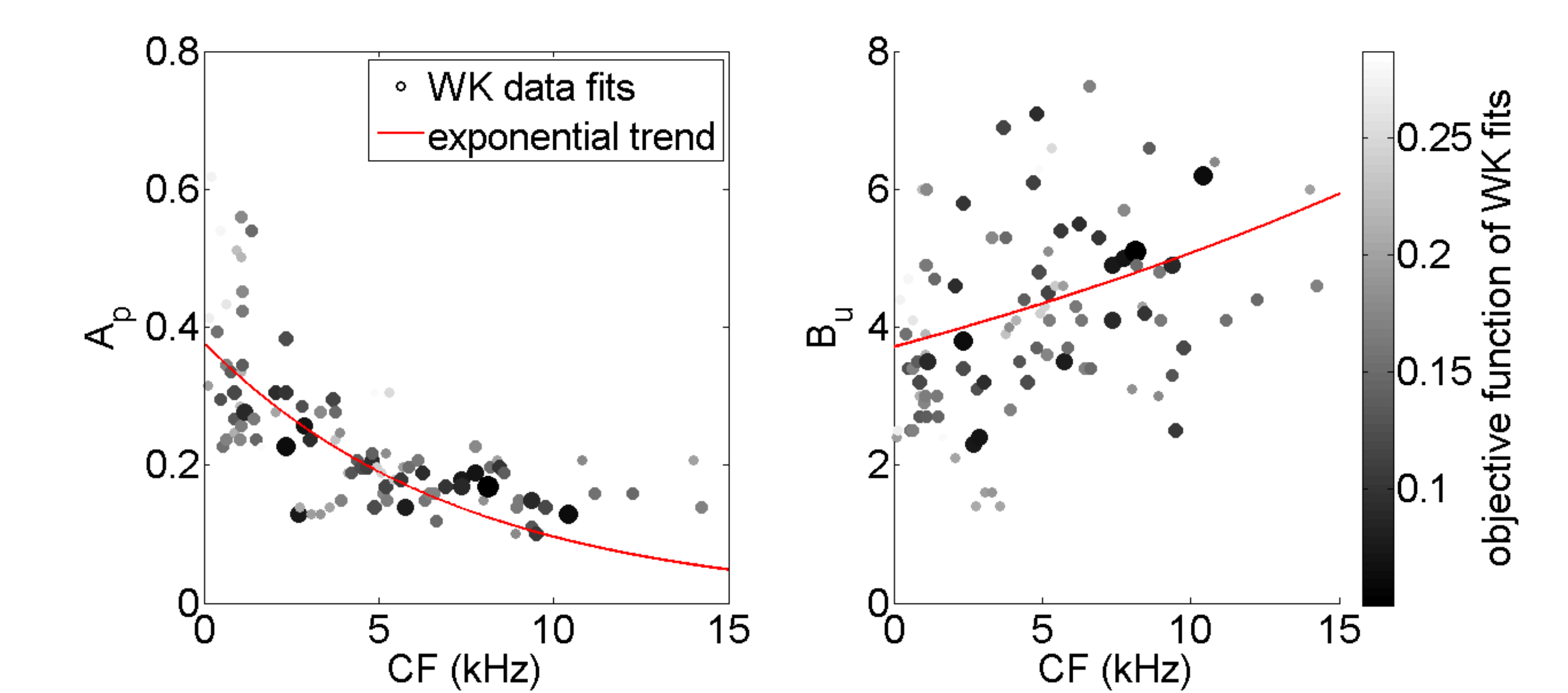}
    \caption{Model constants $\Ap, \Bu$ along the length of the chinchilla cochlea obtained by fitting velocity expression to data and the exponential trend: The model expressions for velocity response is fit to measured data from Wiener Kernels (WK) of chinchilla neural data from points along the length of the cochlea from the base to the apex \cite{ruggeroWK}.  The model constant $\bp$ is fixed at $\bp = 1$  and the estimated model constant values for $\Ap, \Bu$ are plotted as a function of characteristic frequency. The darker shading and larger size of the circle correspond to smaller objective function and hence more reliable estimates. Exponential trend lines (red solid - color online) were obtained from the parameter estimates. The exponential fit functions are in the main text.}
    \label{fig:trends}
\end{figure}

In order to determine the model constants along the length of the cochlea, we use Wiener Kernel data gathered from multiple nerve fibers from the base and apex  (the apical basal transition for chinchilla is at 2.5kHz) \cite{ruggeroWK}. Our estimated values for the model constants are shown in Fig.     \ref{fig:trends}, along with exponential fits of the trend.

The exponential fits, $g(\text{CF}(x)) = a \expn{b \text{CF}(x)}$, yield the following coefficients with 95\% confidence bounds for $\Ap(\text{CF}(x)), \Bu(\text{CF}(x))$.

\begin{itemize}
    \item $a_{\Ap} =      0.3768 \qquad (0.3468, 0.4067)$
    \item $       b_{\Ap} =     -0.1366 \qquad \text{kHz}^{-1} \qquad(-0.1595, -0.1137)$  
    \item $a_{\Bu} =       3.714  \qquad (3.335, 4.093)$
    \item $       b_{\Bu} =     0.03123  \qquad \text{kHz}^{-1} \qquad (0.0153, 0.04715)$
\end{itemize}

As can be seen from the coefficients above, and Fig. \ref{fig:trends}, the trend is more reliable for $\Ap$ than $\Bu$. Note that, while deriving our model expressions, we have assumed that the model constants $\Ap, \Bu, \bp$ do not vary with location (such that the wavenumber is purely a function of $\B$). This assumption still holds provided that the model constants vary slowly, which is quite valid at local scales. We provide support for this extrapolation to spatially varying model constants in appendix \ref{s:spatialvalidity}.

\begin{figure}[htbp!]
    \centering
    \includegraphics[trim={0 0 35cm 0},clip=true, scale=0.22]{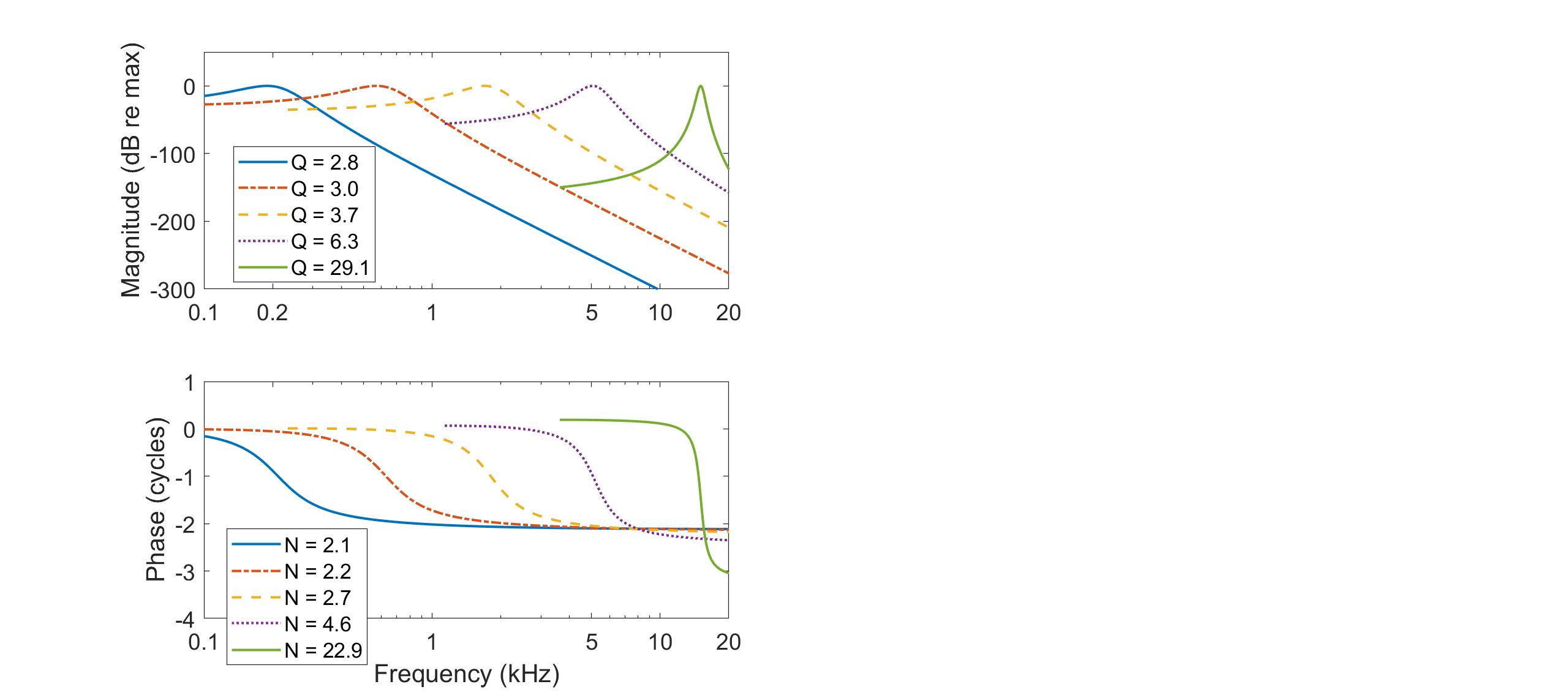}
    \caption{Normalized velocity at various locations using the variable model constants (the 5-parameter) version of the model (color online). The model expressions for velocity response are used to generate velocity responses normalized to their peak absolute value at various locations along the cochlea. The values of the model constants used ($a_{\Ap}, b_{\Ap}, a_{\Bu}, b_{\Bu}, \bp$) are described in section \ref{s:trends}. In the figure, the expressions were extrapolated to non-shortwave regions and beyond the peak. The legends include computed values for the dimensionless quality factor, $Q$, derived from the equivalent rectangular bandwidth, and the normalized center frequency group delay, $N$ in cycles. These are comparable to empirical estimates  \cite{sheratriangle}. The ratio $\frac{Q}{N}$ for the model velocity at these locations is  $1.27-1.38$. For comparison, the average empirical estimate is $1.25$ \cite{sheratriangle}.}
    \label{fig:multiVelocityTFs}
\end{figure}

Using the aforementioned values for model constants, we illustrate normalized OoC velocity at a few locations along the length of the cochlea (Fig \ref{fig:multiVelocityTFs}). Note that only the previously mentioned model constants ($\Ap, \Bu, \bp$, or, $a_{\Ap}, b_{\Ap}, a_{\Bu}, b_{\Bu}, \bp$) are needed to determine the macromechanical responses normalized to their peak, and normalized wavenumber and impedance $\mathcal{P}, \mathcal{V}, k\frac{l}{\B}, \frac{Z}{\rho l \text{CF}(x)}$. However, the absolute forms of some of these variables (and their inter-relations) may require the following constants readily available in the literature for various mammalian species including chinchilla and humans: scalae fluid density $\rho$ which is approximately the density of water; the cochlear map space constant, $l$; and the peak frequency in the very base, $\text{CF}(0)$. For chinchilla, $l = 3.8$ mm, $\text{CF}(0) = 20$ kHz \cite{greenwood}. 

\section{Discussion}
\label{s:discussion}

We previously mentioned that the model may be used for designing auditory filters,  estimating and studying cochlear mechanisms, and intermediate applications. Our discussion of the model features and implementation here \added{primarily} focuses on aspects directly relevant for auditory filter applications. \added{The discussion supports the model and demonstrates the utility of the model features for engineering applications. Many of these features are also particularly important for scientific study as is briefly mentioned at the end of this section.}

\subsection{Relation to Other Models}

\subsubsection{Brief Comparative Notes}

In comparison to models purely for filter-design - e.g. \cite{sumner, zilany, lyoncascade}, to first approximation  \footnote{where we assume scaling symmetry across the modeled region of the cochlea.}, our model requires \replaced{parametrization}{parameterization} of three model constants for all filters (as opposed to parameterizing each filter), while allowing for inherent spatial variation of response characteristics \footnote{Further refinements to spatial variation of response characteristics can be made by allowing for spatial variation of our model parameters - e.g. by using the five model constant version}. The spatial variation of response characteristics is distinctive feature of the cochlea that must be reproduced for filter design.  The inherent spatial variation is possible because of our assumption of scaling symmetry and working in the domain of the transformed independent variable, $\B$. We have shown that our assumption of scaling symmetry of $k$ is appropriate on local scales, as the model constants vary slowly with location - see section \ref{s:model}. For better fits, we have represented the variation of the model constants with slowly varying expressions that retain its validity.

In comparison to mechanistic models - e.g. \cite{zweig76, zweig, meaud,deboer}, our expressions directly extend to \emph{both} the transfer function and traveling wave domain representations (unlike current similar models that are only native to the traveling wave domain)\replaced{and hence}{. Hence, our model} can directly be used to extract and provide information on both fronts \added{(which would be quite desirable for scientific study)} and \added{also} be used for designing \added{both} auditory \replaced{filters}{filter banks and filter cascades}. 

\deleted[moved the rest of this section down and modified it]{ }

\subsubsection{Similarities to Other Models}
\label{s:similaritiesOtherModels}

\begin{figure}
    \centering
    \includegraphics[scale=0.11]{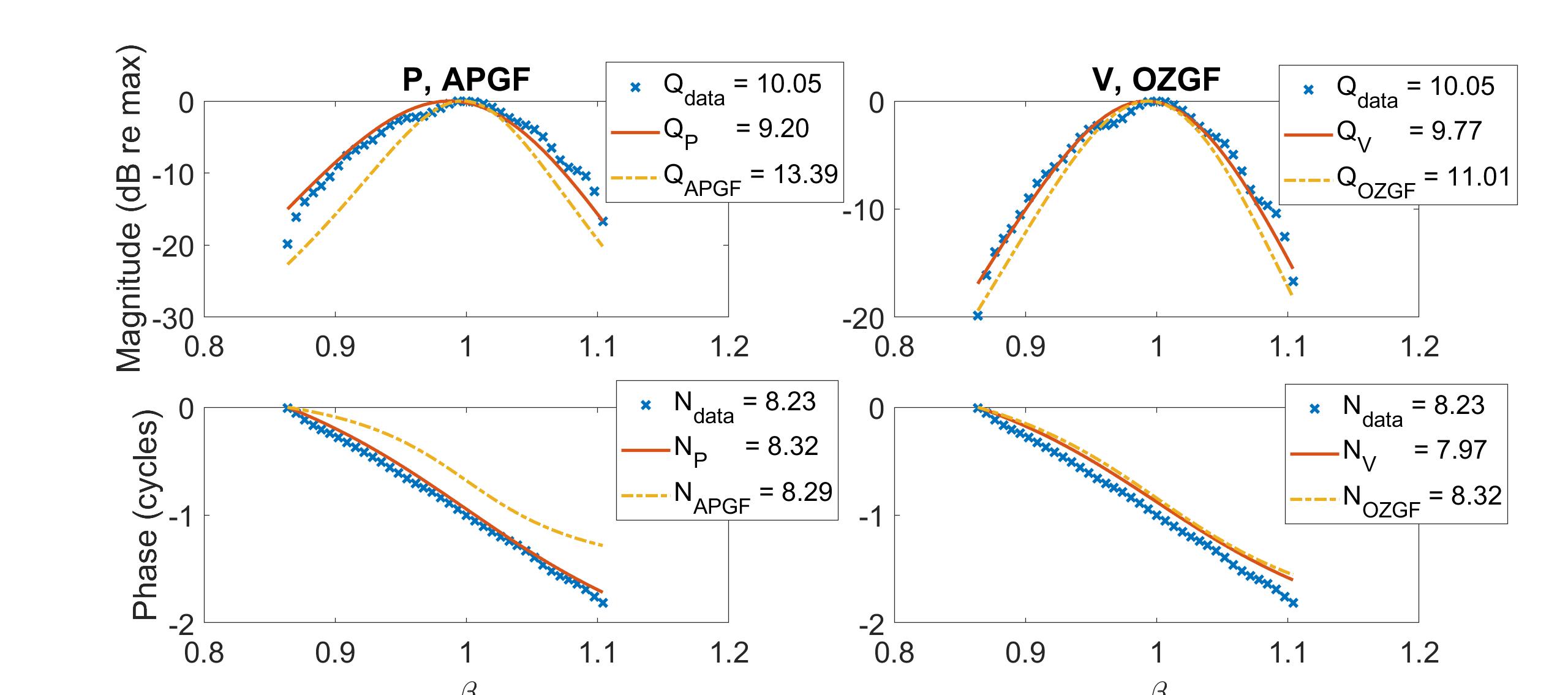}
    \caption{Similarity between model $\mathcal{P}, \mathcal{V}$ form and filters with existing implementations - normalized APGF and OZGF: Our model expression (red line - color online)  for pressure (left) and velocity (right) responses and the APGF (left) and OZGF (right) expressions (yellow dash-dotted - color online) are fit to measured data (blue crosses) from Wiener Kernels of chinchilla neural data in the region of the peak collected at a point where the characteristic frequency is 9.625 kHz  \cite{ruggeroWK}. The magnitude (top) and phase (bottom) are plotted as a function of normalized frequency $\B$. The model constant $\bp$ is fixed at $\bp = 1$ for both $\mathcal{P}, \mathcal{V}$ fits. The imaginary part of the poles APGF and OZGF are equivalently constrained to $2\pi\textrm{CF}(x)$. For the $\mathcal{P}$ fit, the estimated model constant values for $\Ap$, and $\Bu$ are 0.16, and 8.3 respectively, and the objective function value is 0.16. For APGF, $a,b$ (in kHz) and $n$ are $9.68, 61$, and 4, and the objective function value is 0.2. For $\mathcal{V}$, $\Ap$, and $\Bu$ are 0.15, and 6.3, and the objective function value is 0.07. For OZGF, the parameters, $a,b,c$ (in kHz) and $n$ are $14, 61, 185$ and $6$ and the objective function value is 0.11. The legends include computed values for the dimensionless quality factor, $Q$, derived from the equivalent rectangular bandwidth, and the normalized center frequency group delay, $N$ in cycles. }
    \label{fig:comparison}
\end{figure}

Our model's pressure and velocity expressions have similarities to existing decompositions of the  gammatone filter: APGF  and OZGF \cite{Lyon96theall-pole,lyon5} - note that \added{, as they fall purely under the auditory filter category, } these decompositions do not have expressions for wavenumber and impedance. The expression for the APGF, originally derived by finding an approximation for an algorithmic implementation for the original gammatone filter \cite{Lyon96theall-pole}, is $\frac{b^n}{(\tilde{s}^2 +a\tilde{s} + b)^n}$ where $\tilde{s} = i\w$ and $a,b$ are real parametric constants as can be inferred from Eq. 5 of \cite{lyon5} and are chosen for \textit{each} filter. The APGF expression is quite similar to our expression for pressure - Eq. \ref{eq:P} (that only has repeated pairs of complex conjugate poles). The similarity between these two expressions holds if our $C$ is determined entirely from the other parameters, $C \bigg(2\pi \textrm{CF}(x)\bigg)^{2\Bu} = b^n$, and only for integer ($n$) values of $\Bu$. The expression for the OZGF, is $\frac{b^{n-0.5}(\tilde{s}+c)}{(\tilde{s}^2 + a\tilde{s} + b)^n}$ as can be inferred from Eq. 7 of \cite{lyon5}. The similarity of this expression to our $V$ - Eq. \ref{eq:velocity}, which has a real zero in addition to repeated pairs of complex conjugate poles, holds if our zero is not imposed to be the negative of the real part of our pole, our $C$ is determined entirely from other parameters $\frac{i C\Bu}{\rho l \textrm{CF}(x)}\bigg(2\pi \textrm{CF}(x)\bigg)^{2\Bu+1} = b^{n-0.5}$ and only for integer ($n-1$) values of $\Bu$. Figure \ref{fig:comparison} \footnote{To put the APGF and OZGF parameters in terms closer to our own model parameters: For APGF, the negative of the real part of the pole, and $n$ are $2\pi\textrm{CF}(x)0.08$, and 4. For OZGF, the negative of the real part of the pole, the real zero ($c$), and $n$ are $2\pi\textrm{CF}(x)0.12$,  $2\pi\textrm{CF}(x) 3.05$, and $6$. Recall that the imaginary part of the poles APGF and OZGF are  constrained to $2\pi\textrm{CF}(x)$. Note that the factor $2\pi\textrm{CF}(x)$ converts poles and zeros from the dimensionless $s=i\B$ domain to the $\stilde =i \w$ domain.} illustrates the similarity in response form between our $\mathcal{P}$ and APGF, and between our $\mathcal{V}$ and OZGF at a single location \footnote{Improving response fits is not the primary goal of this manuscript. However, if the researcher is only interested in the response variables in our model, and given sufficient data, it would be of interest to compare the form of these models more basal the peak. While we may compare the forms of these models, it is inappropriate to make comparisons regarding the performance of these models based on the fits to data at the peak. This is the case because of lack of data outside of the peak region and potential dependence on datasets \cite{patterson2003} and their sources (e.g. neural or mechanical from various membranes in the organ of Corti) - the models must be general enough to fit various datasets but specific enough to provide good fits to data using a small number of parameters. Additionally, it is inappropriate to compare the goodness of fits of the models due to differences in the number of parameters and/or parameter spaces (e.g. restrictions to integer numbers): For our $\mathcal{P}$, and $\mathcal{V}$, we optimize over two real model constants ($\Ap$, $\Bu$). For APGF, we optimize over a real constant (the real part of the pole) and an integer constant, $n$. For OZGF, we optimize over two real constants (the real part of the pole and the real zero) and one integer, $n$. In all models, we constraint the value of the imaginary part of the pole to correspond to the peak frequency (or peak $\B$).}.

As a result of the aforementioned similarities between our model expressions for $P,V$ and APGF and OZGF, we can leverage current software and hardware implementations for applying our model for auditory filter design purposes \added{. For example, we can benefit from the architectures for OZGF which is particularly amenable to analog implementations \cite{katsiam6} \footnote{This is appropriate for those implementations of our model that are similar to filter banks.}}. This should reduce the additional implementation efforts associated with adopting new models. \deleted{However, we must note that the APGF and OZGF models are in a single independent variable $\w$, and the parameters are chosen for \emph{each} location, $x$. This is as opposed to encoding (and benefiting from) an \textit{implicit} spatial variation as is the case for our model - which reduces the total number of model constants to 2-5 constants in our model (the variation is encoded in our model structure rather than the model constants)}. Due to \replaced{such}{formulation and parameterization} differences, and depending on the implementation scheme used, some degree of modifications will be required in order to leverage APGF and OZGF implementations/algorithms for our model, but the fundamentals of implementation should be similar.

Our model's wavenumber and impedance have similarities to a recent Zweig 2015 model \cite{zweig} which is a 4-parameter model of the mechanistic model type that is centered around the impedance. Therefore we may extrapolate the fundamentals of Zweig's effective formulation for the OoC dynamic representation of force balance to our model.


The similarities between our model and those mentioned above - the APGF, OZGF, and the Zweig model, provide strong support for the expressions of these models and our own. This is especially true as: (1) the derivation methods are fundamentally different than ours and yet arrive at similar expressions for subsets of our variables, and (2) each of the models we have discussed is similar to our model in a different variable ($P, V$, and $Z$).

\subsection{Implementation}
\label{s:implementation}


\subsubsection{Model variable for auditory filter design}
For auditory filter implementations, the appropriate model variable may either be normalized velocity, $\mathcal{V}$, or normalized pressure, $\mathcal{P}$. It is important to note that, for most values of model constants, $\mathcal{V} \approx \mathcal{P}$ near the peak, and hence we may consider both to be representative of a macromechanical cochlear response. Therefore, \added{for purely engineering-type applications}, the major determining factor for using pressure or velocity is simplicity of implementation - this choice is therefore tied to the choice of the implementation scheme. As mentioned earlier, the macromechanical cochlear responses provide good approximations for auditory nerve responses in the peak region \cite{ruggero}, and hence the model can be used for determining neural responses as well. \added{For some scientific purposes that are concerned with relations between variables, it may be desirable to consider pressure and velocity as two separate response variables and pursue a full implementation of the model}.


For many signal-processing applications, a constant quality factor across $\text{CF}(x)$ is desirable (i.e. the bandwidth is proportional to characteristic frequency), just like a wavelet transform. This criterion is simply achievable by using the expressions with constant $\Ap, \Bu, \bp$. In general, to first approximation, or if the frequency range of interest is limited, we suggest using  $\mathcal{V}$ or $\mathcal{P}$ expressions, with constants values for the model constants $\Ap, \Bu, \bp$. When needed, the spatially varying versions, $\Ap(x), \Bu(x), \bp(x)$, may be used. The constant parameterized version of the model has three constants, $\Ap, \Bu, \bp$, whereas the spatially varying parameterized version has five constants, $a_{\Ap}, b_{\Ap}, a_{\Bu}, b_{\Bu}, \bp$.

\subsubsection{Method}


Because the model has closed-form frequency expressions as well as physical underpinnings, it can be implemented in a number of ways, as listed below. This flexibility in implementation is a strength of the model as it can leverage existing efforts for \textit{both} filter bank or filter cascade models that target improved algorithmic or analog circuit efficiencies. In the following, note that $x$ can be thought of as a proxy for filter center frequency, $\text{CF}(x)$. We note that certain implementations are more straightforward if the estimated $\Bu$ is an integer, or rounded to the nearest integer. This rounding yields responses that are closest to that of the exact model near the peak.

\begin{enumerate}
    \item Using the physics-based ordinary differential equation (Eq. \ref{eq:kxPclassical}) along with the closed-form expression for $k$: This is related to an analog implementation of filter cascades due to its relation to transmission lines. By introducing relatively insignificant reverse traveling waves, we can deal with an approximate second order ODE (see, e.g., \cite{deboerreverse}). We then may implement the model using a transmission line which is well suited for time domain analog circuit implementations. The transmission line series impedances are inductors (due to the scalae fluid density) and its shunt impedances must be designed based on our model impedance $Z$ \footnote{Due to the imaginary gain constant (and the effective negative damping), the circuit for $Z$ is not a simple circuit. We expect that there may be additional longitudinal coupling through the shunt impedance of the transmission line in addition to a power source.}. The input to the transmission line is the velocity of the stapes. This implementation is physically somewhat  different than what we have described in this paper, as it replaces $C$ of Eq. \ref{eq:velocity} with $C(\w)$ which is determined by setting the basal boundary of the short-wave region to be the stapes velocity \footnote{\label{note1}This yields $V(x, \w) = \vstapes \frac{1}{s_o} 
    \bigg( \frac{s+\Ap}{\so+\Ap} \bigg) 
    \bigg( \frac{(\so -p)(\so - \pconj)}{(s-p)(s-\pconj)}\bigg)^{\Bu + 1}$ where $\so \triangleq i\frac{\w}{\wmax}$. The specific boundary condition used is $\frac{\partial{P(x, \w)}}{\partial{x}} |_{x=o} = 2i\w\r\vstapes$ where $\r$ is the scala fluid density and $x=0$ is where the stapes is and the short-wave region of the cochlea starts.}, as opposed to some unknown long-wave region between the short-wave region and the stapes which lead to our expressions with the unknown constant, $C$. Our analysis has shown that both assumptions regarding boundary conditions yield similar results for normalized pressure and velocity except closest to the stapes - see appendix \ref{s:choiceofBC}.
    \item Using the integral form of Eq. \ref{eq:kxPclassical} along with the closed-form expression for $k$: This is related to frequency domain digital implementations of filter cascades due to the ability to covert its discrete counterpart into a multiplication of digital filters. Specifically, due to the integral (or sum) in the exponent, the model can be implemented by multiplying subfilters and the spectrum of the stapes velocity.  Note that if the pressure is chosen as the digital filter variable, and the model constants are assumed to be non-spatially varying, then the implementation simply becomes a series of the same repeated filter, in which case it is more convenient to simply used the model closed-form expression for pressure.
    \item Using the closed-form expression for velocity (Eq. \ref{eq:velocity}) or pressure: This is related to frequency domain digital implementations of filter banks, as only a single filter needs to be utilized for each band. The provided expressions must be multiplied by the stapes velocity spectrum. In the case of scaling symmetry where pressure is chosen as the digital filter variable, the model constants are assumed to be non-spatially varying, and the model is implemented with independent variable, $\B$, then only a single filter is needed for \textit{all} bands. This is quite convenient, though adds a degree of complexity due to converting back from $\B$ to frequency and space (or center frequency). 
\end{enumerate}

\begin{figure}[htbp!]
    \centering
    \includegraphics[scale = 0.3]{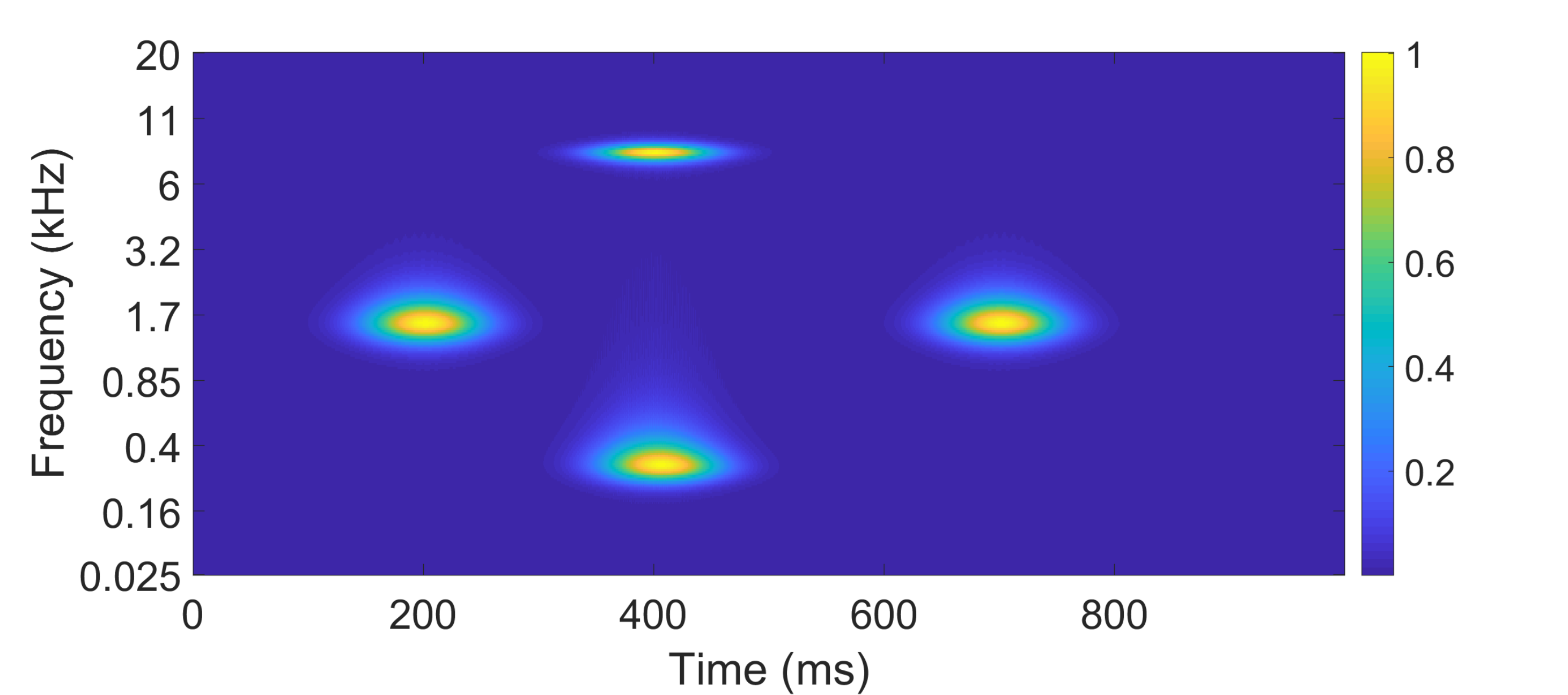}
    \caption{Simulated model response to complicated input to illustrate implementability: The figure (color online) illustrates that the model can be used to process complicated signals. The input consists of short tone bursts and has the following form $\vstapes = \sum_{i=0}^3 e^{-(t-t_i)^2/T^2} sin(2\pi f_i t)$, where $T=50, f_0 = 1.5, t_0 = 200, f_1 = 8, t_1 = 400, f_2 = 1.5, t_2 = 700, f_3 = 0.3, t_3 = 400$ with frequency in kHz, and time in ms. To generate the response along the length of the cochlea in this figure, the 5-parameter version of the model was used with model constants values in section \ref{s:trends}. Note that the 3-parameter version of the model can alternatively be used particularly if constant quality factors (\textit{not} to be confused with constant bandwidths) are desirable. The velocity response was determined (Eq. \ref{s:velocity}), from the frequency domain equivalent of the input, then transformed into the time domain response whose envelope is shown in the figure.} 
    \label{fig:chirp}
\end{figure}

A future direction is to pursue the analog and digital implementations outlined above.  In addition to their utility as model tests, Fig. \ref{fig:velocityNeural} as well as those in appendix \ref{s:morefits} illustrate the third (closed-form $V$) implementation. As another example, we demonstrate the suitability of using the digital implementation for determining the cochlear response at various locations to complicated signals - see Fig. \ref{fig:chirp}. As seen in the figure, in response to an input consisting of short tone bursts, the model cochlea moves maximally at locations where the CF$(x)$ is close to the tonal frequency of the tone burst. The \replaced{respond}{response} is limited beyond the durations of the tone bursts. The localization of the cochlear responses in time and space are in agreement with our expectation regarding the cochlea.

\subsection{Linearity Limitations and Extensions}


As mentioned in section \ref{s:physics}, the model is linear and therefore presumably most appropriate for low-level stimuli. \added{Many engineering and scientific applications are addressable using the appropriate linear models}. To incorporate nonlinearity in the model, we suggest three potential paths:

\begin{itemize}
\item \added{Make the model constants level-dependent to make the model quasi-linear: For example, as done in \cite{verhulst}, and the quasi-linear OZGF. We may utilize the similarities between our model's response variable and OZGF to choose and incorporate level-dependence in the parameters.}
\item \added{Link the model to an external component which has nonlinearity - e.g. similar to what was done by \cite{lyoncascadereview}}
\item \replaced{By incorporating}{Incorporate} nonlinearity directly in time domain differential equations of velocity or pressure
\end{itemize}

\subsection{On Model Tests using Data}
\added{We have tested the model with data at multiple points and for, and using, multiple variables. The multivariate and multifaceted tests provide rigor and further support for the model. The multivariate nature of testing is beneficial partly because a test of the model using a single variable is a test of all variables in the model - this is possible because they are tied together by a single set of model constants.


Qualitative tests regarding impedance and processing complex sounds support the form of the model expressions and underlying assumptions. The discussion about boundary condition assumptions (appendix \ref{s:choiceofBC}) also support underlying assumptions we made in order to derive closed-form expressions for response variables from our constructed wavenumber expression. We compared the model with data using multiple quantitative tests centered around the wavenumber and response variables in which we inverted for values of model constants - this was made possible by the simple closed-form expressions and small number of parameters. The tests use both real and imaginary parts - rather than just response amplitude. Tests using response expressions use data from multiple locations - rather than from a single location. Testing the model using information from multiple locations, each with its own objective function (rather than pooled), serves to demonstrate that the model expressions are appropriate beyond a single point or single set of parameter values. We also included values of response characteristics $Q$ and $N$ corresponding the to response fits and found that they are similar to those of physiological data.

Further tests that can be performed to further support the model and determine its representational limits include using currently available data:  mechanical measurements (presumably more directly related to our model $P,V$ than neural approximations to mechanical variable), data from different species, or psychoacoustic rather than physiological data. It is also  desirable to use existing simultaneous measurements (e.g. of variables that can be approximated as $P$ and $V$ \cite{dong}) in order to further test the underlying model assumptions. }

\subsection{On Model Features}
\added{The features of our developed model have been been introduced in previous sections. Here, we exemplify a few benefits of our model features for various readers interested primarily in specific applications. Recall that our model has closed-form expressions for all mechanistic and response/filter variables that are parameterized and related through a single small set of model constants, primarily dependent on a single independent variable, and has a physical basis.


Naturally, none of these features matter if a model does not provide an appropriate representation for the applications of interest. Our model tests using multiple variables illustrate the model's appropriateness in fitting data. The resultant response characteristics, $Q, N$, are also comparable to those computed from data.

\subsubsection{Features for Intuition and Analysis}
The simple closed-form expressions with a small number of parameters and a single independent variable are easy to derive insights from. They allow for building intuition then determining the entire model along with deep model-behavior analyses. This intuition is not only about the behavior of a variable but also facilitates gaining intuition regarding how the behavior of various model variables depends on each other or on the model constants. Building intuition in model development and the ease of studying the model is primarily relevant for those interested in filter analysis and cross-filter analysis  - e.g. \cite{lyon5}. Building intuition about how the various model variables behave relative to each other allows for gaining intuition desirable for scientific study - for example, the behavior of the pressure amplitude is a window into the behavior of the imaginary part of the wavenumber behavior and real part of the impedance.

\subsubsection{Features for Parameterizability}
The simplicity of the model expressions allows for estimating parameter values based on data or characteristics \footnote{Note that while we have used a classical fitting scheme for determining the model constants (which is most appropriate for model testing), the model constants can also, in fact, be estimated from response characteristics (such as bandwidth, and group delay) as apparent from the form of the closed-form expressions. The resultant estimated values can then be used to determine the model mechanistic and response / filter variables}. This parameterizability is key to enabling further pursuits in any application. Parameterizability, ease of determining parameter values (and therefore any of the model variables) from desired values of responses or response characteristics such as quality factors and group delays or from other model behavior - is a much appreciated feature in engineering spheres, and one that can facilitate scientific studies aiming to infer mechanistic information from experimental data or their characteristics. Parametizablility is further facilitated by the fact that the model variables are all closed-form expressions tied together by a single set of model constants. Beyond utilizing model parameterizability for engineering purposes and scientific study, parameterizability also makes testing tractable \footnote{Note that current parametric mechanistic models in the literature cannot be tested without fixing the values of a number of model constants.}. We have found that parameterizability is not generally achieved in current mechanistic models and cascade auditory filters, though it is a valued feature of simple auditory filter banks.

\subsubsection{Features for Flexibility of Implementation}
The model has representations for multiple variables ($k,Z,P,V$), closed-form expressions and physical inter-relations between its variables. All three of these model properties result in flexibility in implementing the model for auditory filter design purposes, \added{and for scientific study}. This fact, including model implementability as both filter bank and filter cascade, is discussed in detail in section \ref{s:implementation} of the manuscript. Flexibility of implementation is particularly desirable for engineers. To our knowledge, no current appropriate models allow for such a degree of implementational flexibility.

\subsubsection{Features for Efficiency}
The closed-form expressions for the model variables are simple and require a small number of model constants. These features translate into \replaced{intuitive design and implementational efficiency (generally, closed-form expressions are the most computationally efficient)}{ implementational efficiency (generally, closed-form expressions are the most computationally efficient). }.

\subsubsection{Features for Non-Classical Applications}
The model is derived using a physical-phenomenological approach, it contains simple closed-form expressions for both mechanistic and filter / response variables parameterized by the same set of parameter values, and it ties together the transfer function and traveling wave perspectives. These dualities in the model and its development enable exploring new possible applications that require or utilize bridging between current perspectives. To our knowledge, no current model contains both mechanistic and filter components with appropriate features to do so.}

\section{Conclusion}

We developed a linear model of the mammalian cochlea that is primarily valid in the short-wave (peak) region, which is of primary importance in transmitting signals to the brain, and consequently, in designing auditory filters and for scientific study. \added{The model features make it appropriate as a single framework for either application and for bridging between them}. We developed the model using a mixed physical-phenomenological approach and introduced the model closed-form expressions for the wavenumber\added{, impedance, pressure,} and velocity in terms of three model constants, and a single independent variable, $\B$ that relates space and frequency. 

The simplicity of the model and its small number of parameters contribute to \added{its intuitive and analyzable nature, and }its efficiency \added{in implementation and inverting for parameter values (parameterizability)}. We tested the model using real and imaginary parts of chinchilla data and have shown fits for multiple variables. Using data from multiple variables furthers our confidence in testing the model \deleted[clarifying and expanding upon notes regarding testing in a new section]{(note that the internal consistency of the model is already insured by the physical equations which are appropriate approximations in the peak region)}. We have also shown comparisons to computed response quality factors and group delays. The model predicts impedances that are qualitatively consistent with current literature. Furthermore, we provide values for the model constants as a function of location in the chinchilla that can be used for (1) scientific study of the differences between the base and the apex, and (2) engineering contributions that require constructing auditory filters centered around various characteristic frequencies.  We demonstrated the model's utility for simulating cochlear motion in responses to complicated signals. We showed that the model formulation links auditory responses to the underlying mechanisms through a single set of model constants that allows for determining one variable from another. We have also shown that the model offers flexibility in terms of filter bank and filter cascade formulations and analog and digital implementation.

Most other models do not have closed form expressions for all variables (wavenumber, impedance, pressure and velocity) in the frequency domain and it is not simple to derive one variable from another. The model encodes the aforementioned variables as (1) closed-form expressions, that are (2) a function of the same set of three (or five) model constants. Therefore, \added{the model forms a single framework for both mechanistic and response (or filter) variables and can be used to bridge between scientists and engineers. Furthermore, }the model can be used to determine  any one of these variables from another, as may be relevant for scientific study -  e.g. determining variation in negative damping and amplification as derived from the auditory responses for various species and cochlear regions. We discussed various implementations of the models to pursue analog and digital schemes for \replaced{auditory filter applications}{response-centric implementations}, and illustrated two such implementations.


\begin{acknowledgments}
The authors would like to thank John Guinan, Dennis Freeman, John Rosowski, and Alan Grodzinsky for helpful discussions, and two anonymous reviewers for useful comments. 
Support by the Harvard-MIT Speech and Hearing Bioscience \& Technology (SHBT) Program and Grant Nos. T32 DC00038 (SHBT) and R01 DC003687 (CAS) from the NIDCD.
\end{acknowledgments}

\appendix 

\section{Method for Fitting Model to Data}
\label{s:fitmethod}

Here, we explain the method by which we fit the model to velocity data in sections \ref{s:velocity}, \ref{s:trends}, and appendix \ref{s:morefits}. We also use a similar method for fitting the wavenumber in section \ref{s:wavenumber}.

Before fitting the model to the Wiener Kernel data, we chose data sets and processed them as follows: We chose data sets that did not have too many ripples (visually); We eliminated data points below a noise threshold of $-20$ dB; We eliminated data points outside the range of $\B=0.5-1.5$.

We performed the latter two due to the nature of the neural experiments that generated Wiener Kernel data \cite{ruggero}: the confidence in the measured Wiener Kernel data is much higher close to the peak.

We obtain the estimates for model constants $\Ap, \Bu$ by finding the minimum of an objective function on an $\Ap$ x $\Bu$ grid while fixing $\bp = 1$. It is appropriate to fix $\bp$ since the CF of the single points from which the datasets were collected can be easily approximated from the measured magnitude curves \footnote{Small deviations from $\bp=1$ are consistent with the model, and would naturally improve the fit if that is the primary concern. However, we limit ourselves to the smallest possible number of parameters here (two).}. The simple brute force approach we employ avoids issues of local minima. We define the objective function to be $\sqrt{\frac{\sum_{i=1}^m |\Delta|^2}{m}}$ where $m$ is the number of data points, and $\Delta$ is the complex residual, $\mathcal{V}_{data}-\mathcal{V}_{model}$. Note that this form for the objective function weighs the real and imaginary parts equally, and gives greater weight to regions where $\mathcal{V}$ is large and hence emphasizes the peak region. The $\Ap$ x $\Bu$ grid is constructed such that each of $\Ap$ and $\Bu$ has a range of 100 logarithmically-spaced values from 0.1 to 10 \footnote{\aftertbme{Note that if the only interest was in obtaining the best fit for either $k$, or $V$, then the model constants need not be constrained to be positive real numbers as we have chosen for our 2D search grid. However, in doing so, the model would no longer satisfy physical constraints 1, and 2.}}.  The grid values were chosen such that they incorporate expected model constant values.





\section{Additional Fits}
\label{s:morefits}

\begin{figure}[htbp]
    \centering
    \includegraphics[scale=0.12]{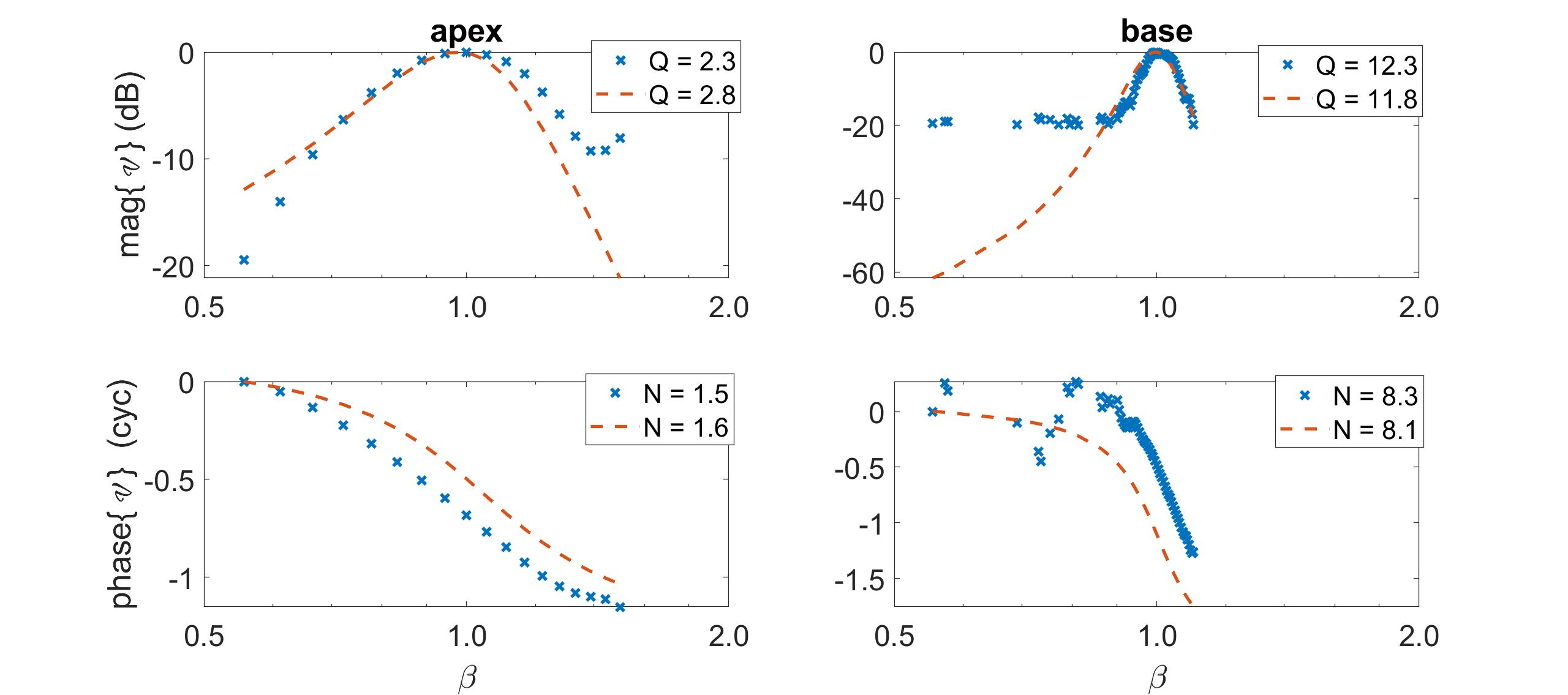}
    \caption[Model test using velocity expressions II]{Model test using velocity expression: The model (red dashed - color online) expression for velocity response is fit to measured data (blue crosses) from Wiener Kernels of chinchilla neural data in the region of the peak \cite{ruggeroWK}. The fits are for data from the apex collected at a point where the characteristic frequency is 140 Hz (left) and for data from the base collected at a point where the characteristic frequency is 14 kHz (right). The magnitude (top) and phase (bottom) are plotted as a function of normalized frequency $\B$. The model constant $\bp$ is fixed at $\bp = 1$ for both fits. For the fit to the point in the apex, the estimated model constant values for $\Ap$, and $\Bu$ are 0.3, and 2.4 respectively, and the objective function value is 0.19. For the fit to the point in the base, the estimated model constant values for $\Ap$, and $\Bu$ are 0.1, and 4.1 respectively, and the objective function value is 0.18. The legends include computed values for the dimensionless quality factor, $Q$, derived from the equivalent rectangular bandwidth, and the normalized center frequency group delay, $N$ in cycles.}
    \label{fig:velocityNeural2}
\end{figure}

\begin{figure}[htbp!]
    \centering
    \includegraphics[scale = 0.12]{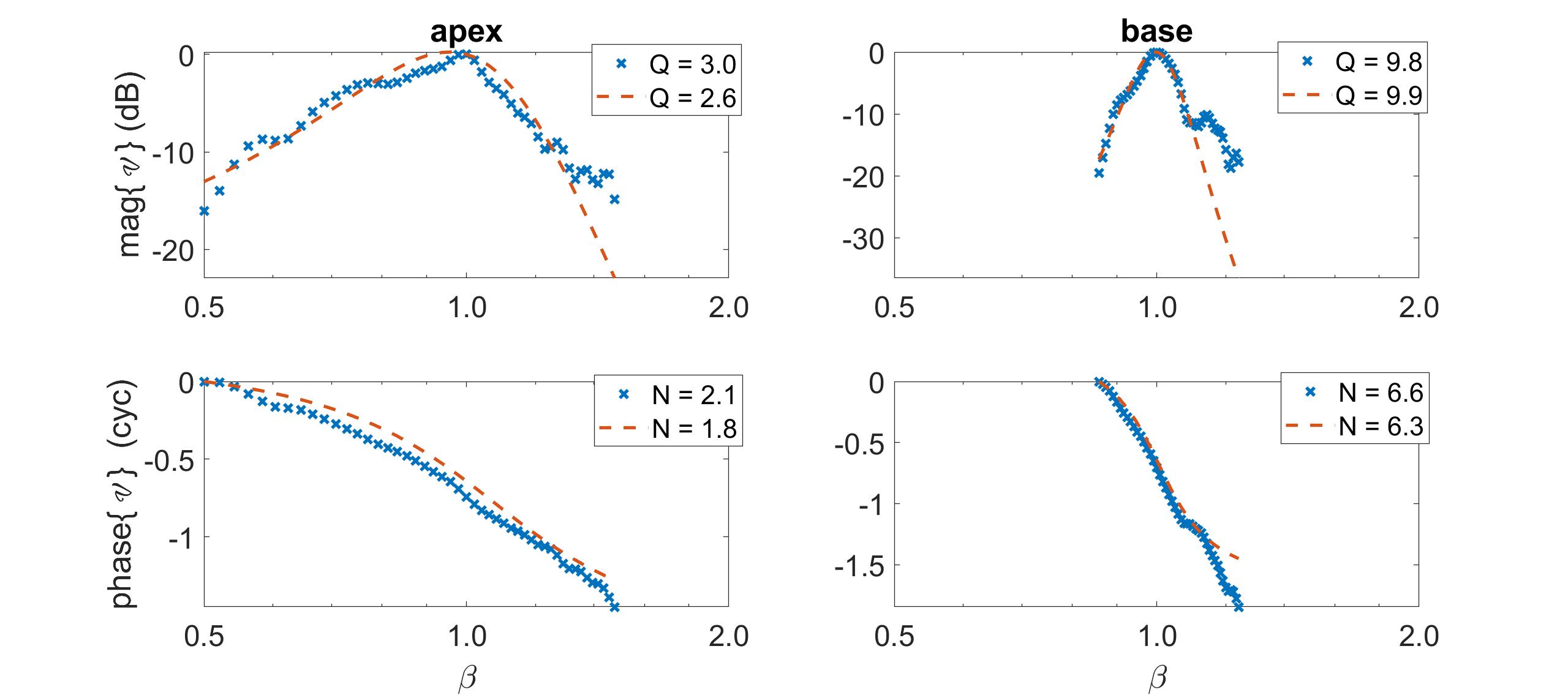}
    \caption[Model test using velocity expressions III]{Model test using velocity expression: The model (red dashed - color online) expression for velocity response is fit to measured data (blue crosses) from Wiener Kernels of chinchilla neural data in the region of the peak \cite{ruggeroWK}. The fits are for data from the apex collected at a point where the characteristic frequency is 1.5 kHz (left) and for data from the base collected at a point where the characteristic frequency is 7.5 kHz (right). The magnitude (top) and phase (bottom) are plotted as a function of normalized frequency $\B$. The model constant $\bp$ is fixed at $\bp = 1$ for both fits. For the fit to the point in the apex, the estimated model constant values for $\Ap$, and $\Bu$ are 0.4, and 3.4 respectively, and the objective function value is 0.12. For the fit to the point in the base, the estimated model constant values for $\Ap$, and $\Bu$ are 0.1, and 3.4 respectively, and the objective function value is 0.12. The legends include computed values for the dimensionless quality factor, $Q$, derived from the equivalent rectangular bandwidth, and the normalized center frequency group delay, $N$ in cycles.}
    \label{fig:velocityNeural3}
\end{figure}

\begin{figure}[htbp!]
    \centering
    \includegraphics[scale = 0.12]{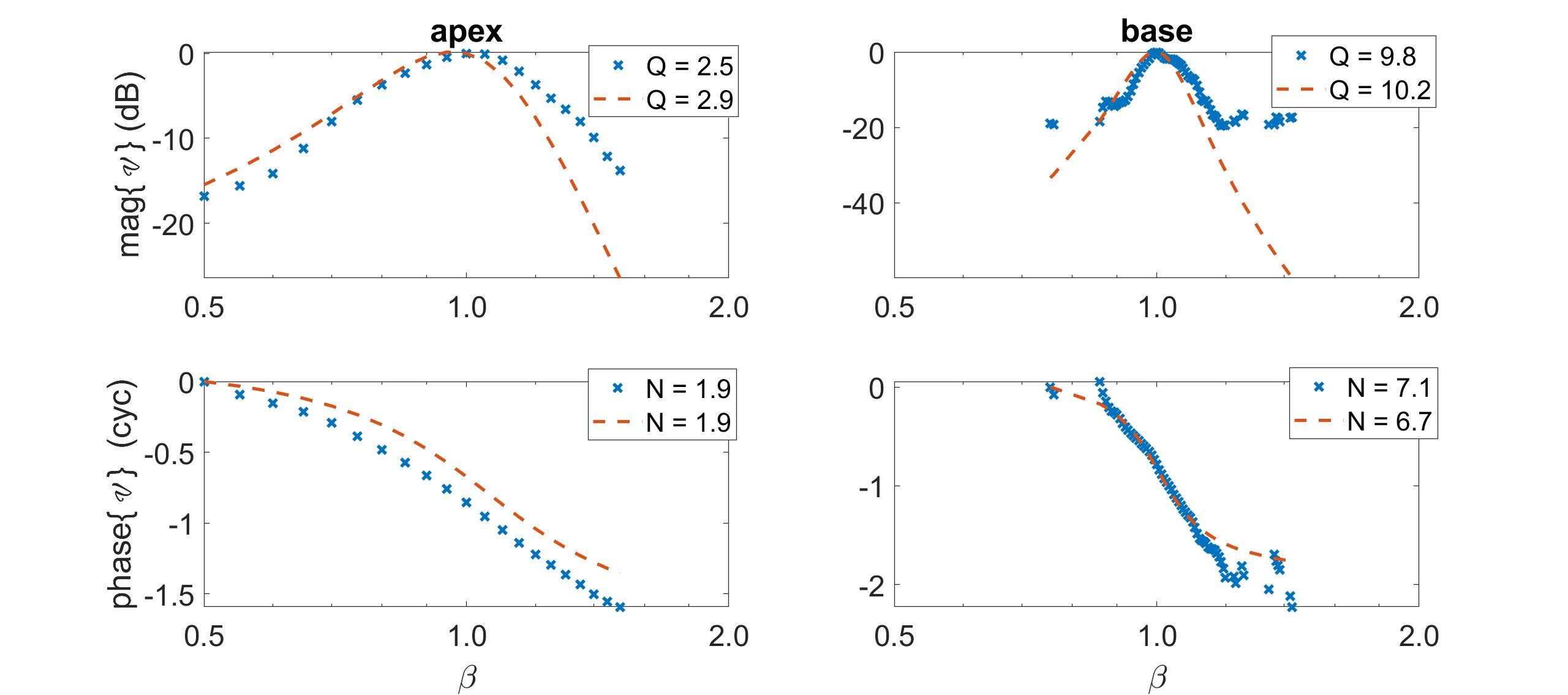}
    \caption[Model test using velocity expressions IV]{Model test using velocity expression: The model (red dashed - color online) expression for velocity response is fit to measured data (blue crosses) from Wiener Kernels of chinchilla neural data in the region of the peak \cite{ruggeroWK}. The fits are for data from the apex collected at a point where the characteristic frequency is 625 Hz (left) and for data from the base collected at a point where the characteristic frequency is 9.4 kHz (right). The magnitude (top) and phase (bottom) are plotted as a function of normalized frequency $\B$. The model constant $\bp$ is fixed at $\bp = 1$ for both fits. For the fit to the point in the apex, the estimated model constant values for $\Ap$, and $\Bu$ are 0.35, and 3.4 respectively, and the objective function value is 0.17. For the fit to the point in the base, the estimated model constant values for $\Ap$, and $\Bu$ are 0.1, and 3.4 respectively, and the objective function value is 0.17. The legends include computed values for the dimensionless quality factor, $Q$, derived from the equivalent rectangular bandwidth, and the normalized center frequency group delay, $N$ in cycles.}
    \label{fig:velocityNeural4}
\end{figure}

In this appendix, we give additional examples of our model velocity fit to neural Wiener Kernel data  \cite{ruggero} - see Figs. \ref{fig:velocityNeural2}-\ref{fig:velocityNeural4}. The figures show better fits of magnitude and phase near $\B=1$, and show that the model captures the slope of the phase. The figures also show the model fits in cases where the data has multiple ripples (Fig. \ref{fig:velocityNeural3}), or is unreliable outside of the peak region (Fig. \ref{fig:velocityNeural2}, and the base case of Fig. \ref{fig:velocityNeural4}). The apex case of Fig. \ref{fig:velocityNeural4} illustrates a case where the model magnitude deviates from the data close to the peak.


\section{Validity of Extrapolating Model by Incorporating Spatial ~ Variability in Model Constants} 
\label{s:spatialvalidity}

\begin{figure}[htbp!]
    \centering
    \includegraphics[scale = 0.3]{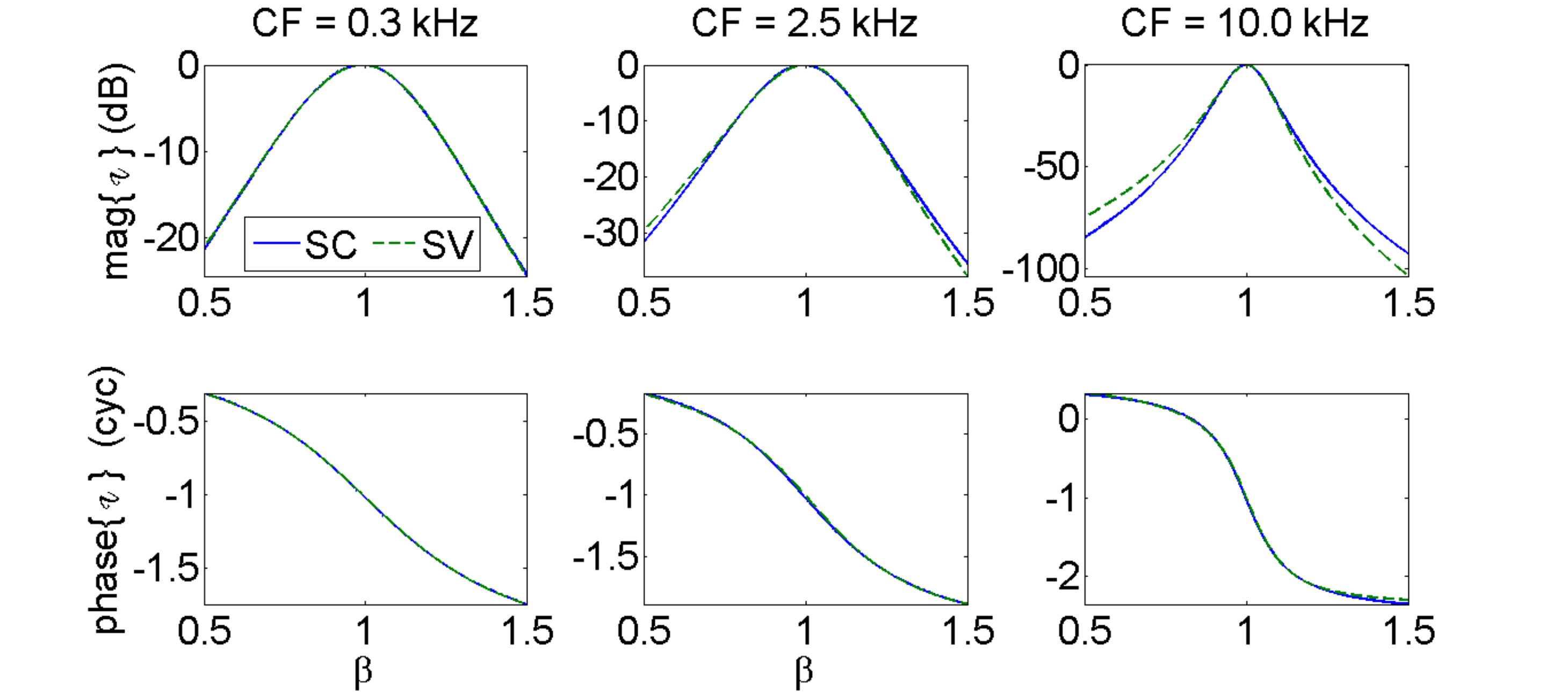}
    \caption{Validity of local wavenumber scaling symmetry assumption: The figure shows the magnitude and phase of the normalized velocity computed by numerical integration from the wavenumber expression for three locations along the length of the chinchilla cochlea. Two sets of model constants where used: SV (green dashed - color online) allows for spatial variation of the model constants according to section \ref{s:trends}; SC (blue solid), assumes that the model constants do not vary with space in the local region spanning $\B = 0.5-2.0$ close to the peak. In this case, the model constants take on the \textit{single} values prescribed by their CF according to section \ref{s:trends}.The figure shows that the two cases yield similar results.}
    \label{fig:robustnessspatialvariation}
\end{figure}

Figure \ref{fig:robustnessspatialvariation} shows the normalized velocity at three locations along the length of the chinchilla cochlea with (1) spatially constant model constants, $\Ap, \Bu, \bp$ chosen for each of the three locations individually, and (2) spatially varying model constants. The similarity between these two cases supports the idea that the model constants are effectively slowly varying and supports our assumption of local wavenumber scaling symmetry.

In order to compare the two cases, the velocity was numerically generated by integrating the wavenumber. This is appropriate as, based on our analysis, we have assumed that the choice of boundary condition does not have much effect on the response near its peak. The integration is over a fixed range of $\B$, rather then $x$, and hence the velocity response in the apical region is not that influenced by the characteristics in the base.



\section{Choice of Basal Boundary Condition}
\label{s:choiceofBC}

The objective of this appendix is \textit{not} to formulate expressions for the long-wave region, but rather to justify how we handle the effects of the long-wave region on the short-wave model. We construct our expressions for the short-wave region (close to where the peak in responses occurs) as that is where the majority of signal filtering occurs in the active cochlea, and hence is our region of interest. 

Between the stapes and the short-wave region, the long-wave approximation to the 2D wave equation mostly holds. From the perspective of the short-wave region, the long-wave region acts as a boundary condition, and hence contributes a frequency dependent integration constant for Eq. \ref{eq:kxPclassical} in determining the pressure from the wavenumber.

As mentioned in section \ref{s:velocity}, our expressions for the model responses, assume that this frequency dependent factor in the pressure, $C(\w)$ is a frequency independent factor, $C$ (which we treat as an unknown). In other words, we assume that the effect of the long-wave contribution on the short-wave model alters the amplitude and phase of the responses in an approximately frequency-independent manner (i.e. it does not contribute much to tuning). To study the effect of varying boundary conditions, we generated responses for $P$, $V$ as expressed in the main text, and for an extreme case where $C(\w)$ is determined by setting the basal boundary of the short-wave region to be the stapes. Our analysis has shown that both assumptions regarding boundary conditions yield similar results for normalized pressure and velocity except closest to the stapes \footref{note1}. This supports our assumption that the choice of boundary condition does not have a significant effect on the model, and by extension, the effects of incorporating long-wave region effects into the short-wave model is relatively insignificant (except closest to the stapes).






\end{document}